\documentclass[dvipdfmx]{ptephy_v1}

\preprintnumber{XXXX-XXXX} 

\begin{document}

\title{Causal production of the electromagnetic energy flux and role of the negative energies in Blandford-Znajek process}


\author[1,2]{Kenji Toma}
\author[3]{Fumio Takahara}

\affil[1]{Frontier Research Institute for Interdisciplinary Sciences, Tohoku University, Sendai 980-8578, Japan \email{toma@astr.tohoku.ac.jp}}
\affil[2]{Astronomical Institute, Tohoku University, Sendai 980-8578, Japan}
\affil[3]{Department of Earth and Space Science, Graduate School of Science, Osaka University, Toyonaka 560-0043, Japan}



\begin{abstract}%
  Blandford-Znajek process, the steady electromagnetic energy extraction from a rotating black hole (BH), is widely believed to work for driving relativistic jets in active galactic nuclei, gamma-ray bursts and Galactic microquasars, although it is still under debate how the Poynting flux is causally produced and how the rotational energy of the BH is reduced. We generically discuss the Kerr BH magnetosphere filled with a collisionless plasma screening the electric field along the magnetic field, extending the arguments of Komissarov and our previous paper, and propose a new picture for resolving the issues. For the magnetic field lines threading the equatorial plane in the ergosphere, we find that the inflow of particles with negative energy as measured in the coordinate basis is generated near that plane as a feedback from the Poynting flux production, which appears to be a similar process to the mechanical Penrose process. For the field lines threading the event horizon, we first show that the concept of the steady inflow of negative electromagnetic energy is not physically essential, partly because the sign of the electromagnetic energy density depends on the coordinates. Then we build an analytical toy model of a time-dependent process both in the Boyer-Lindquist and Kerr-Schild coordinate systems in which the force-free plasma injected continuously is filling a vacuum, and suggest that the structure of the steady outward Poynting flux is causally constructed by the displacement current and the cross-field current at the in-going boundary between the plasma and the vacuum. In the steady state, the Poynting flux is maintained without any electromagnetic source.
\end{abstract}

\subjectindex{xxxx, xxx}

\maketitle

\section{Introduction}
\label{sec:intro}

The driving mechanism of collimated outflows or jets with relativistic speeds which are observed in active galactic nuclei (AGNs), gamma-ray bursts, and Galactic microquasars is one of the major problems in astrophysics. A most widely discussed model is based on Blandford-Znajek (BZ) process, the electromagnetic energy extraction from a rotating black hole (BH) along magnetic field lines threading it \citep{blandford77}. This process produces Poynting-dominated outflows, which may be collimated by the pressure of the surrounding medium such as the accretion disk and the disk wind \citep[e.g.][]{mckinney06,tchekho11}. The particles in the outflow can be gradually accelerated depending on the geometrical structure \citep[e.g.][]{komi_vlahakis09,lyubarsky09,toma13}, which is consistent with the recent observational implications from the radio jet in the giant elliptical galaxy M87 \citep{asada14,kino15} \citep[see also][]{porth15}.

BZ process was proposed by a pioneering paper of \citep{blandford77}, who found steady, axisymmetric, force-free solutions of Kerr BH magnetosphere in the slow rotation limit where the outward angular momentum (AM) and Poynting fluxes are non-zero along the field lines threading the event horizon. This was followed by demonstrations of analytical and numerical magnetohydrodynamic (MHD) solutions \citep[e.g.][]{beskin00,koide02,komissarov05,mckinney06,tchekho11,mckinney12} and other force-free solutions \citep[e.g.][]{komissarov04,ruiz12,contopoulos13,beskin13}. However, the physical mechanism how the fluxes are created in the electromagnetically-dominated plasma has not been clearly explained. In contrast, the origin of pulsar winds is identified definitely with the rotation of the matter-dominated central star. The rotation velocity of the matters of the star $\mathbf{V}_\varphi$ and the magnetic field threading the star $\mathbf{B}$ provide the electromotive force $\mathbf{V}_\varphi \times \mathbf{B}$ on charges, maintaining the electric field $\mathbf{E}$ and the poloidal electric currents (with the toroidal magnetic field $\mathbf{B}_\varphi$) which form the outward Poynting flux $\mathbf{E} \times \mathbf{B}_\varphi/4\pi$ in the magnetosphere. As its feedback, the rotation of the stellar matters slows down \citep{goldreich69} \citep[see also a review in][hereafter TT14]{toma14}. BZ process, working in the electromagnetically-dominated BH magnetosphere, does not include any matter-dominated region in which the poloidal magnetic field is anchored. One should also note that the toroidal magnetic field cannot be produced in vacuum just by the rotation of the space-time \citep{wald74,komissarov04}. Then how the electric field and currents forming the AM and Poynting fluxes are created and how the BH rotational energy is reduced in BZ process are not simple problems and have been still matters of debate. See recent discussions in \cite{komissarov09} (hereafter K09) and \cite{beskin10}.

Among the numerous calculations, the force-free numerical simulations performed by \cite{komissarov04} (hereafter K04) are most insightful for investigating the essential points on the origin of the fluxes. TT14 extended the arguments in K04 and K09 and analytically showed that for open magnetic field lines threading the ergosphere in the steady, axisymmetric Kerr BH magnetosphere, the situation of no electric potential difference with no poloidal electric current (i.e. no outward AM or Poynting flux) cannot be maintained. The origin of the electric potential differences is ascribed to the ergosphere. It was also shown that for the open field lines threading {\it the equatorial plane in the ergosphere}, the poloidal currents are driven by electric field (perpendicular to the magnetic field) stronger than the magnetic field in the ergosphere, where the force-free condition is violated (see also Section~\ref{sec:equat} below).

In this paper, we mainly discuss the field lines threading the event horizon. Some theorists consider that the membrane paradigm \citep{thorne86} is useful for effectively understanding the production mechanism of the AM and Poynting fluxes for such field lines \citep[e.g.][]{penna13,penna14}. This interprets the condition at the horizon as a boundary condition \citep{znajek77,blandford77} and the horizon as a rotating conductor which creates the potential differences and drives the electric currents in an analogy with the unipolar induction for pulsar winds explained above. However, the horizon does not actively affect its exterior, but just passively absorbs particles and waves \citep{punsly89}. The condition at the horizon should be interpreted as a regularity condition \citep[][K04]{beskin00}. The mechanism of producing the steady AM and Poynting fluxes has to work outside the horizon, making the physical quantities consistent with the regularity condition.

For such a causal flux production, some other theorists proposed a scenario that certain types of negative energies (as measured in the coordinate basis, i.e. as measured at infinity) created outside the horizon flows towards the horizon, resulting in the positive outward energy flux. This is an analogy with the mechanical Penrose process, in which the rotational energy of a BH is extracted by making it absorb negative-energy particles \citep{penrose69,bardeen72,bejger12}. However, MHD simulations demonstrate that no regions of negative particle energy are seen in the steady state \citep{komissarov05}, although a transient inflow of negative particle energy is possible as a feedback from generation of an outward MHD wave \citep{koide02}. The role of {\it negative electromagnetic energy density} in the steady state has been discussed recently \citep[K09;][]{lasota14,koide14}, although the concept of `advection of the steady field' is ambiguous. Below we show that the sign of the electromagnetic energy density depends on the coordinates, and thus the negative field energy is not physically essential (see Section~\ref{sec:negative_EM} below).

We argue that the causal production mechanism of the electromagnetic AM and Poynting fluxes cannot be fully understood by investigating only the steady-state structure. We examine a time-dependent process evolving towards the steady state with an analytical toy model to clarify how the steady outward fluxes are created. In order to find the essential physics, our analysis is performed both in the Boyer-Lindquist (BL) and the Kerr-Schild (KS) coordinate systems. Most of the previous analytical studies used the BL coordinates \citep[e.g.][]{takahashi90,beskin00,menon05,okamoto06,beskin13,pu15} \citep[but see][]{mckinney04}, most of the recent numerical simulations used the KS coordinates \citep[e.g. K04;][]{komissarov05,mckinney06,tchekho11,mckinney12} \citep[but see e.g.][]{beckwith08,contopoulos13}, and both of them focused on the steady-state structure.\footnote{The force-free electrodynamics without decomposition of tensors into spatial and temporal components has also been developed \citep{carter79,uchida97,gralla14,yang15}.} Our new analytical studies of time-dependent process in the BL and KS coordinates will be highly helpful for understanding physics in BZ process.

This paper is organized as follows. In Section~\ref{sec:formulation}, we explain our formulation of general relativistic electrodynamics, set generic assumptions for Kerr BH magnetosphere, and review the recent analytical understandings given by K04 and TT14. Section~\ref{sec:equat} concentrates on the field lines threading the equatorial plane, for which we show the flux production mechanism and the role of the negative energy of particles. In Section~\ref{sec:horizon}, we explain differences between the equatorial plane and the horizon, and then we focus on the field lines threading the horizon, discussing differences of the electromagnetic structures as seen in the BL and KS coordinates and the role of the negative electromagnetic energy density. In Section~\ref{sec:unsteady}, we discuss the time-dependent process towards the steady state. Section~\ref{sec:conclusion} is devoted to conclusion.

\section{Formulation and Assumptions}
\label{sec:formulation}

\subsection{The $3+1$ decomposition of space-time}

The space-time metric can be generally written as
\begin{eqnarray}
  ds^2 = g_{\mu\nu} dx^\mu dx^\nu
  = -\alpha^2 dt^2 + \gamma_{ij} (\beta^i dt + dx^i)(\beta^j dt + dx^j),
  \label{eq:metric}
\end{eqnarray}
where $\alpha$ is called the lapse function, $\beta^i$ the shift vector and $\gamma_{ij}$ the three-dimensional metric tensor of the space-like hypersurfaces. Those hypersurfaces are regarded as the absolute space at different instants of time $t$ \citep[cf.][]{thorne86}. We focus on Kerr space-time with fixed BH mass $M$ and angular momentum $J$. (The electromagnetic field with outward fluxes which we consider below is a test field for Kerr space-time.) We adopt the units of $c=1$ and $GM=1$, for which the gravitational radius $r_g = GM/c^2 = 1$. We use the dimensionless spin parameter $a \equiv J/(M r_g c)$.

Kerr space-time has two symmetries, i.e. $\partial_t g_{\mu\nu} = \partial_\varphi g_{\mu\nu} = 0$. These correspond to the existence of the Killing vector fields $\xi^\mu$ and $\chi^\mu$. In the coordinates $(t, \varphi, r, \theta)$,
\begin{equation}
  \xi^\mu = (1, 0, 0, 0), ~~~\chi^\mu = (0, 1, 0, 0).
\end{equation}
The event horizon, where $g^{rr} = 0$, is located at $r_{\rm H} = 1+ \sqrt{1-a^2}$. The ergosphere is the region $r < r_{\rm es} = 1 + \sqrt{1-a^2\cos^2\theta}$, where the Killing vector $\xi^\mu$ is space-like, $\xi^2 = g_{tt} = -\alpha^2 + \beta^2 > 0$. At infinity, this space-time asymptotes to the flat one.

The local fiducial observer \citep[FIDOs;][]{bardeen72,thorne86}, whose world line is perpendicular to the absolute space, is described by the coordinate four-velocity
\begin{equation}
  n^\mu = \left(\frac{1}{\alpha}, \frac{-\beta^i}{\alpha}\right), ~~~
  n_\mu = g_{\mu\nu} n^\nu = (-\alpha, 0, 0, 0).
\end{equation}
The AM of this observer is $n_\varphi = 0$, and thus FIDO is also a zero AM observer \citep[ZAMO;][]{thorne86}. Note that the FIDO frame is not inertial, but it can be used as a convenient orthonormal basis to investigate the local physics \citep{misner73,thorne86,punsly08}. It should also be confirmed that the FIDOs are time-like, physical observers (i.e. $n^\mu n_\mu = -1$).

In the BL coordinates, one has the following non-zero metric components:
\begin{eqnarray}
  &&\alpha = \sqrt{\frac{\varrho^2 \Delta}{\Sigma}}, ~~ \beta^\varphi = -\frac{2ar}{\Sigma}, \nonumber \\
  && \gamma_{\varphi\varphi} = \frac{\Sigma}{\varrho^2}\sin^2\theta, ~~ \gamma_{rr} = \frac{\varrho^2}{\Delta}, ~~ \gamma_{\theta\theta} = \varrho^2,
\end{eqnarray}
where
\begin{eqnarray}
  \varrho^2 = r^2 + a^2 \cos^2\theta, ~~~~ \Delta = r^2 + a^2 - 2r, ~~~~
  \Sigma = (r^2 + a^2)^2 - a^2 \Delta \sin^2\theta.
\end{eqnarray}
BL FIDOs rotate in the same direction as the BH with the coordinate angular velocity
\begin{equation}
\Omega \equiv \frac{d\varphi}{dt} = -\beta^\varphi = \frac{2ar}{\Sigma}.
\end{equation}
The BL coordinates have the well-known coordinate singularity ($\alpha = 0$ and $\gamma_{rr} = \infty$, where $\Delta = 0$) at the horizon. The BL FIDOs are physical observers only outside the horizon.

The KS coordinates have no coordinate singularity at the event horizon. The coordinates $t$ and $\varphi$ are different from those in the BL coordinates. The non-zero metric components in this coordinate system are:
\begin{eqnarray}
  &&\alpha = \frac{1}{\sqrt{1+z}}, ~~ \beta^r = \frac{z}{1+z}, ~~ \gamma_{r\varphi} = -a (1+z) \sin^2\theta, \nonumber \\
  &&\gamma_{\varphi\varphi} = \frac{\Sigma}{\varrho^2}\sin^2\theta, ~~
  \gamma_{rr} = 1+z, ~~ \gamma_{\theta\theta} = \varrho^2,
\end{eqnarray}
where $z = 2r/\varrho^2$ \citep[K04;][]{mckinney04}. The KS spatial coordinates are no longer orthogonal ($\gamma_{r\varphi} \neq 0$). From the space-time symmetries,
\begin{eqnarray}
  && g_{\mu\nu} \xi^\mu \xi^\nu = g_{tt} = -\alpha^2 + \beta^2, \nonumber \\
  && g_{\mu\nu} \xi^\mu \chi^\nu = g_{t\varphi} = \gamma_{\varphi j} \beta^j = \beta_\varphi, \nonumber \\
  && g_{\mu\nu} \chi^\mu \chi^\nu = g_{\varphi\varphi} = \gamma_{\varphi\varphi}
\end{eqnarray}
are the same in the BL and KS coordinates. We should note that the KS FIDOs are different from the BL FIDOs (K04).

\subsection{The $3+1$ electrodynamics}

We follow the definitions and formulations of K04 for electrodynamics in Kerr space-time (except for keeping $4\pi$ in Maxwell equations), in a similar way to TT14 \citep[see also K09, references therein, and][]{landau75}. The covariant Maxwell equations $\nabla_\nu {}^* F^{\mu\nu} = 0$ and $\nabla_\nu F^{\mu\nu} = 4\pi I^\mu$ are reduced to
\begin{equation}
\nabla \cdot \mathbf{B} = 0, ~~~~ \partial_t \mathbf{B} + \nabla \times \mathbf{E} = 0, 
\label{eq:maxwell1}
\end{equation}
\begin{equation}
\nabla \cdot \mathbf{D} = 4\pi \rho, ~~~~ - \partial_t \mathbf{D} + \nabla \times \mathbf{H} = 4\pi \mathbf{J},
\label{eq:maxwell2}
\end{equation}
where $\nabla \cdot \mathbf{C}$ and $\nabla \times \mathbf{C}$ denote 
$(1/\sqrt{\gamma})\partial_i(\sqrt{\gamma} C^i)$ and $e^{ijk} \partial_j C_k$, respectively, and 
$e^{ijk} = (1/\sqrt{\gamma}) \epsilon^{ijk}$ is the Levi-Civita pseudo-tensor of the absolute space. The condition of zero electric and magnetic susceptibilities for general fully-ionized plasmas leads to the following constitutive equations,
\begin{equation}
\mathbf{E} = \alpha \mathbf{D} + \boldsymbol{\beta} \times \mathbf{B},
\label{eq:relation_E}
\end{equation}
\begin{equation}
\mathbf{H} = \alpha \mathbf{B} - \boldsymbol{\beta} \times \mathbf{D},
\label{eq:relation_H}
\end{equation}
where $\mathbf{C} \times \mathbf{F}$ denotes $e^{ijk} C_j F_k$. At infinity, $\alpha=1$ and $\boldsymbol{\beta} = 0$, so that $\mathbf{E} = \mathbf{D}$ and $\mathbf{H} = \mathbf{B}$. Here $D^\mu = F^{\mu\nu}n_\nu$ and $B^\mu = -{}^*F^{\mu\nu}n_\nu$ are the electric and magnetic fields as measured by the FIDOs, while $E^\mu = \gamma^{\mu\nu} F_{\nu\alpha} \xi^\alpha$ and $H^\mu = -\gamma^{\mu\nu} {}^*F_{\nu\alpha} \xi^\alpha$ are the electric and magnetic fields in the coordinate basis, where $\gamma^{\mu\nu} = g^{\mu\nu} + n^\mu n^\nu$. The current $\mathbf{J}$ is related to the current as measured by the FIDOs, $\mathbf{j}$, as
\begin{equation}
  \mathbf{J} = \alpha\mathbf{j} - \rho\boldsymbol{\beta}.
  \label{eq:relation_j}
\end{equation}
See Appendix~\ref{app:current} on the relation between convective current and particle velocity.

The covariant energy-momentum equation of the electromagnetic field
$\nabla_\nu T^{\nu}_{\mu} = -F_{\mu\nu} I^\nu$ gives us the AM equation as
\begin{equation}
  \partial_t l + \nabla \cdot \mathbf{L} = -(\rho \mathbf{E} + \mathbf{J} \times \mathbf{B})\cdot \mathbf{m},
  \label{eq:AM}
\end{equation}
and the energy equation as
\begin{equation}
  \partial_t e + \nabla \cdot \mathbf{S} = - \mathbf{E} \cdot \mathbf{J},
  \label{eq:energy}
\end{equation}
where $\mathbf{C} \cdot \mathbf{F}$ denotes $C^i F_i$, $\mathbf{m} = \partial_\varphi$,
\begin{equation}
  l = \alpha T_\varphi^t
  = \frac{1}{4\pi}(\mathbf{D}\times \mathbf{B})\cdot \mathbf{m}
\end{equation}
is the AM density,
\begin{equation}
  \mathbf{L} = -(\mathbf{E}\cdot \mathbf{m}) \mathbf{D} - 
(\mathbf{H}\cdot \mathbf{m}) \mathbf{B}
+ \frac{1}{2} (\mathbf{E}\cdot \mathbf{D} + \mathbf{B}\cdot \mathbf{H})\mathbf{m}
\end{equation}
is the AM flux ($L^i = \alpha T^i_\varphi$),
\begin{equation}
  e = -\alpha T_t^t
  = \frac{1}{8\pi} (\mathbf{E} \cdot \mathbf{D} + \mathbf{B} \cdot \mathbf{H})
\end{equation}
is the energy density, and
\begin{equation}
  \mathbf{S} = \frac{1}{4\pi}\mathbf{E} \times \mathbf{H}
\end{equation}
is the Poynting flux ($S^i = -\alpha T^i_t$).

\subsection{Kerr BH magnetosphere}
\label{subsec:magnetosphere}

\subsubsection{Electromagnetic fields}
\label{sec:setup}

We study the axisymmetric electromagnetic field in Kerr space-time which is filled with a plasma. (The steadiness of the field is partly discarded in Section~\ref{sec:unsteady}.) We put the additional assumptions similarly to TT14: (1) The poloidal $\mathbf{B}$ field produced by the external currents (whose distribution is symmetric with respect to the equatorial plane) is threading the ergosphere. We call the field lines threading the ergosphere `ergospheric field lines'. (2) The plasma in the BH magnetosphere is dilute and collisionless, but its number density is high enough to screen the electric field along the $\mathbf{B}$ field lines, i.e.
\begin{equation}
\mathbf{D} \cdot \mathbf{B} = 0.
\end{equation}
The energy density of the particles is much smaller than that of the electromagnetic fields. (3) The gravitational force is negligible compared with the Lorentz force. (The gravitational force overwhelms the Lorentz force in a region very close to the event horizon \citep{punsly08}, but the physical condition in that region hardly affects its exterior.)

The condition $\mathbf{D} \cdot \mathbf{B} = 0$ and equation (\ref{eq:relation_E}) lead to $\mathbf{E} \cdot \mathbf{B} = 0$. {\it In the steady state,} we have $\nabla \times \mathbf{E} = 0$, which means that $\mathbf{E}$ is a potential field, and the axisymmetry leads to $E_\varphi = 0$. Then one can write
\begin{equation}
  \mathbf{E} = -\boldsymbol{\omega} \times \mathbf{B}, ~~~ \boldsymbol{\omega} = \Omega_{\rm F} \mathbf{m}.
  \label{eq:omegaF}
\end{equation}
Substituting this equation into $\nabla \times \mathbf{E} = 0$, one obtains
\begin{equation}
  B^i \partial_i \Omega_{\rm F} = 0.
\label{eq:Omega_F_const}
\end{equation}
That is, $\Omega_{\rm F}$ is constant along each $\mathbf{B}$ field line. The $\mathbf{E}$ field is also described by $E_i = -\Omega_{\rm F} \partial_i \Psi$ in terms of the magnetic flux function $\Psi$, so that each $\mathbf{B}$ field line is equipotential and $\Omega_{\rm F}$ corresponds to the potential difference between the field lines.

In the steady, axisymmetric state, the equations (\ref{eq:AM}) and (\ref{eq:energy}) are reduced to
\begin{equation}
\nabla \cdot \left(\frac{-H_\varphi}{4\pi} \mathbf{B}_{\rm p} \right) =
B^i \partial_i \left(\frac{-H_\varphi}{4\pi} \right) = - (\mathbf{J}_{\rm p} \times \mathbf{B}_{\rm p}) \cdot \mathbf{m},
\label{eq:AM2}
\end{equation}
\begin{equation}
\nabla \cdot \left(\Omega_{\rm F} \frac{-H_\varphi}{4\pi} \mathbf{B}_{\rm p} \right) = B^i \partial_i \left(\Omega_{\rm F} \frac{-H_\varphi}{4\pi} \right) = -\mathbf{E} \cdot \mathbf{J}_{\rm p},
\label{eq:energy2}
\end{equation}
where the subscript p denotes the poloidal component. Here one sees that the poloidal AM and Poynting fluxes are described by
\begin{equation}
  \mathbf{L}_{\rm p} = \frac{-H_\varphi}{4\pi} \mathbf{B}_{\rm p}, ~~~~~
  \mathbf{S}_{\rm p} = \Omega_{\rm F} \frac{-H_\varphi}{4\pi} \mathbf{B}_{\rm p},
\end{equation}
respectively. It should be noted that $H_\varphi = {}^*F_{\mu\nu} \xi^\mu \chi^\nu$ and $\Omega_{\rm F} = -F_{t\theta}/F_{\varphi\theta}$ are the same in the BL and KS coordinates (K04).

TT14 shows that the condition $\Omega_{\rm F} > 0$ is inevitable for the ergospheric field lines in the steady, axisymmetric state (see also K04; K09). Furthermore, for the ergospheric field lines crossing the outer light surface (see Section~\ref{sec:basic_motion}), the condition
\begin{equation}
  \Omega_{\rm F} > 0, ~~~ H_\varphi \neq 0
\end{equation}
has to be maintained, i.e. the poloidal AM and Poynting fluxes are steadily non-zero (TT14). The following discussion in this paper focuses on how their values are causally determined and the role of the negative energies as measured in the coordinate basis.

\subsubsection{Particle motions and light surfaces}
\label{sec:basic_motion}

Under the assumption (2) for the magnetospheric plasma stated in Section~\ref{sec:setup}, the force-free condition $\rho\mathbf{E} + \mathbf{J} \times \mathbf{B} = 0$ (or $\rho \mathbf{D} + \mathbf{j} \times \mathbf{B} = 0$) is satisfied when $D^2 < B^2$ (e.g. K04; TT14; see also Appendix~\ref{app:current}).\footnote{Finite particle mass may cause some inertial drift currents to flow across the field lines, which transfer the AM and energy between the particles and the electromagnetic fields \citep{kuijpers15}. We assume that this effect is negligible for the flux production.} Then equation (\ref{eq:AM2}) indicates
\begin{equation}
  B^i \partial_i H_\varphi = 0 ~~~({\rm for}~ D^2 < B^2).
  \label{eq:forcefree}
\end{equation}
Equations (\ref{eq:AM2}) and (\ref{eq:energy2}) mean that no AM or energy is exchanged between the particles and the electromagnetic fields.

In this case, in the BL coordinates, the particles drift in the azimuthal direction with angular velocity $\Omega_{\rm F}$ when $B_\varphi = 0$ (TT14). The light surfaces are thus defined as where the four-velocity of a particle with the coordinate angular velocity $\Omega_{\rm F}$ becomes null, i.e. $f(\Omega_{\rm F}, r, \theta) = 0$, where
\begin{equation}
  f(\Omega_{\rm F}, r, \theta) \equiv (\xi + \Omega_{\rm F} \chi)^2 =
  -\frac{\varrho^2\Delta}{\Sigma} + \gamma_{\varphi\varphi} (\Omega_{\rm F} - \Omega)^2.
  \label{eq:f}
\end{equation}
There can be two light surfaces; the outer light surface (outside which $f>0$) and the inner light surface (inside which $f>0$). In the case of $0<\Omega_{\rm F}<\Omega_{\rm H}$, $\Omega_{\rm F} = \Omega + \sqrt{\varrho^2 \Delta/\Sigma \gamma_{\varphi\varphi}} > \Omega$ at the outer light surface, and $\Omega_{\rm F} = \Omega - \sqrt{\varrho^2 \Delta/\Sigma \gamma_{\varphi\varphi}} < \Omega$ at the inner light surface, which is located in the ergosphere \citep[][K04; TT14]{takahashi90}. The condition $0 < \Omega_{\rm F} < \Omega_{\rm H}$ is satisfied when the outward Poynting flux is non-zero either for the field lines threading the horizon \citep{blandford77} or the ergospheric field lines threading the equatorial plane (TT14). Note that $f(\Omega_{\rm F}, r, \theta)$ is a scalar, so that the location of each light surface is the same in the BL and KS coordinates.

If $D^2 > B^2$ is realized, the cross-field current flows, i.e. $\mathbf{J}_{\rm p} \times \mathbf{B}_{\rm p} \neq 0$ (TT14). The force-free condition is violated, and $H_\varphi$ varies along a field line.

\section{Field lines threading the equatorial plane}
\label{sec:equat}

\subsection{Production of AM and Poynting fluxes}

The mechanism of $\Omega_{\rm F}$ and $H_\varphi$ being determined along the ergospheric field lines threading the equatorial plane has been already clarified (K04; TT14). Generally in the BL coordinates, one has from equations (\ref{eq:relation_E}), (\ref{eq:relation_H}), and (\ref{eq:omegaF})
\begin{equation}
(B^2 - D^2)\alpha^2 = -B^2 f(\Omega_{\rm F}, r, \theta) + \frac{1}{\alpha^2} (\Omega_{\rm F} - \Omega)^2 H_\varphi^2.
\end{equation}
The key point is that $H_\varphi = 0$ on the equatorial plane because of the symmetry. Therefore, the region of $f(\Omega_{\rm F}, r, \theta) > 0$ (i.e. inside the inner light surface, which is within the ergosphere) can satisfy the condition $D^2 > B^2$ around the equatorial plane, driving the poloidal current to flow across the field lines. (Note that $B^2 - D^2 = F_{\mu\nu} F^{\mu\nu}/2$ is a scalar, so that one has $D^2 > B^2$ also in the KS coordinates.) This leads to $H_\varphi \neq 0$ outside the region where $D^2 > B^2$, which we call `current crossing region'. The value of $\Omega_{\rm F}$ will be regulated so that the current crossing region is finite (see Figure 4 of TT14), and thus it is expected to depend on the microphysics in the ergosphere. The values of $\Omega_{\rm F}$ and $H_\varphi$ will be determined by the conditions around the equatorial plane and at infinity.

In the current crossing region, $\mathbf{D}$ is in the opposite direction of $\mathbf{E}$, i.e. $\mathbf{D} \cdot \mathbf{E} < 0$, as seen in the BL coordinates (see Figure 3 of TT14). This leads to $(\mathbf{J}_{\rm p} \times \mathbf{B}_{\rm p}) \cdot \mathbf{m} < 0$ and $\mathbf{E} \cdot \mathbf{J}_{\rm p} < 0$, which generate the poloidal electromagnetic AM and Poynting fluxes (see equations \ref{eq:AM2} and \ref{eq:energy2}). (We confirm that $\mathbf{D} \cdot \mathbf{E} < 0$ also in the KS coordinates in Appendix~\ref{app:B2D2}.)

All the ergospheric field lines crossing the outer light surface have $\Omega_{\rm F} > 0$ and $H_\varphi < 0$ for the northern hemisphere ($H_\varphi > 0$ for the southern hemisphere), while `the last ergospheric field line', which passes the equatorial plane at $r = r_{\rm es}$, has $\Omega_{\rm F} = H_\varphi = 0$. This means that the current flows inward along the ergospheric field lines and outward along the last ergospheric field line (see Figure~\ref{fig:summary} below). Correspondingly, the current crossing region extends over $r_{\rm H} < r < r_{\rm es}$. Such a poloidal current structure prevents the BH from charging up.

\subsection{Production of particle negative energy}
\label{sec:equa_eom}

Equations (\ref{eq:AM2}) and (\ref{eq:energy2}) imply that the particles in the current crossing region lose their AMs and energies by the feedback, $+(\mathbf{J}_{\rm p} \times \mathbf{B}_{\rm p}) \cdot \mathbf{m}$ and $+\mathbf{E}\cdot \mathbf{J}_{\rm p}$, from the production of the electromagnetic AM and Poynting fluxes. We find that this feedback can make the particles have negative energy as measured in the coordinate basis.

\begin{figure}[t]
\centering\includegraphics[scale=0.5]{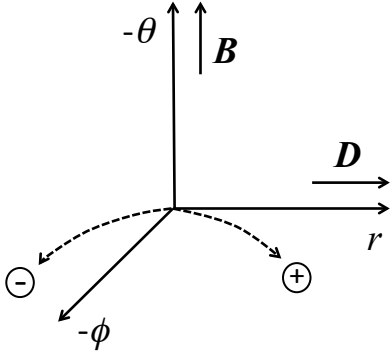}
\caption{
Motion of the positively (negatively) charged particle near the equatorial plane in the BL coordinates. This schematic picture is applicable both in the BL coordinate basis and in the BL FIDO orthonormal basis.
}
\label{fig:motion}
\end{figure}

The production of the particle negative energy can be explained by showing the particle motions in the local orthonormal basis carried with the BL FIDOs, in which the equation of a particle motion with four-velocity $\mathbf{u}$, three-velocity $\mathbf{v}$, charge $q$, and mass $m$ is written as
\begin{equation}
\frac{d\hat{u}_i}{d\hat{t}} = \frac{q}{m}(\hat{D}_i + \epsilon_{ijk} \hat{v}^j \hat{B}^k),
\end{equation}
where $\hat{C}_i$ denotes the vector component in respect of the FIDO's orthonormal basis \citep[][TT14]{thorne86}. In this basis one can investigate local, instantaneous particle motions under the Lorentz force as special relativistic dynamics. (The FIDO frame is not inertial and a particle feels the gravitational force, although we neglect it compared with the Lorentz force, based on the assumption (3) set in Section~\ref{sec:setup}.) The AM and energy per mass of a particle as measured in the coordinate basis are
\begin{eqnarray}
  l_{\rm p} &=& u_\mu \chi^\mu = \gamma_{\varphi\varphi} (v^\varphi - \Omega) u^t
  = \sqrt{\gamma_{\varphi\varphi}} \hat{v}^\varphi \hat{u}^t,
  \label{eq:AM_p} \\
  e_{\rm p} &=& -u_\mu \xi^\mu = [\alpha^2 + \gamma_{\varphi\varphi} \Omega (v^\varphi - \Omega)] u^t
  = (\alpha + \sqrt{\gamma_{\varphi\varphi}}\Omega \hat{v}^\varphi) \hat{u}^t,
  \label{eq:energy_p}
\end{eqnarray}
where we have used $\hat{v}^\varphi = (\sqrt{\gamma_{\varphi\varphi}}/\alpha)(v^\varphi - \Omega)$ \citep[cf.][]{punsly08}.

\begin{figure}[t]
\centering\includegraphics[scale=0.8]{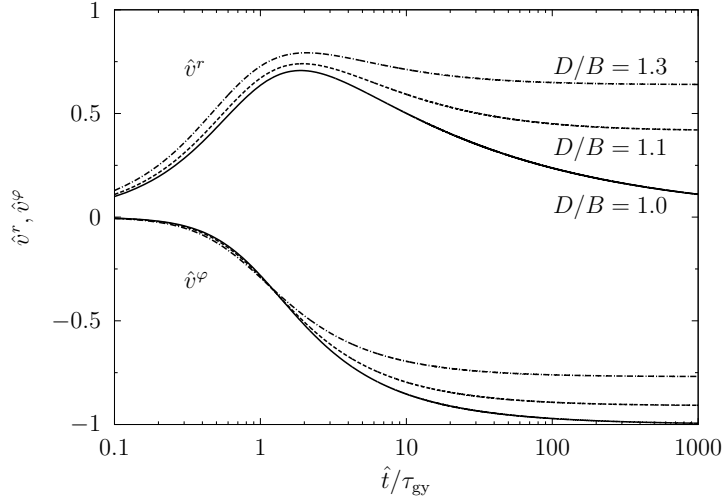}
\caption{
Velocity components $\hat{v}^r$ (upper three lines with positive values) and $\hat{v}^\varphi$ (lower three lines with negative values) of the posively charged particle in the fixed BL FIDO's orthonormal basis as functions of time normalized by gyration time scale $\tau_{\rm gy} = m/q|\hat{\mathbf{B}}|$. The solid, dashed, and dot-dashed lines are calculation results for $|\hat{\mathbf{D}}|/|\hat{\mathbf{B}}| = 1.0, 1.1,$ and $1.3$, respectively. The initial conditions are $\hat{v}^r = \hat{v}^\varphi = 0$.
}
\label{fig:motcal}
\end{figure}

Near the equatorial plane, the $\hat{\mathbf{B}}$ field is approximately perpendicular to that plane, because $B_\varphi = H_\varphi/\alpha = 0$ at that plane, and then the $\hat{\mathbf{D}}$ field is radial in that plane (see Figure~\ref{fig:motion}). The motion of a test particle can be easily solved in such fields \citep{landau75}.  For the case of $D^2 \ge B^2$ which we focus on, the positively (negatively) charged particles are accelerated in the directions of $\hat{\mathbf{D}}$ ($-\hat{\mathbf{D}}$) and $\hat{\mathbf{D}} \times \hat{\mathbf{B}}$. In Figure~\ref{fig:motcal}, we show the calculation results for $|\hat{\mathbf{D}}|/|{\hat{\mathbf{B}}}| = 1.0, 1.1,$ and $1.3$, where we fix the basis and assume that the electromagnetic fields are uniform. For $D^2 = B^2$ (i.e. $|\hat{\mathbf{D}}|/|\hat{\mathbf{B}}| = 1.0$) in particular, the particles are strongly accelerated in the direction of $\hat{\mathbf{D}} \times \hat{\mathbf{B}}$, and then one obtains
\begin{equation}
  \hat{v}^\varphi \approx -1
\end{equation}
in several tens of gyro radius scale $\ell_{\rm gy} = m/q|\hat{\mathbf{B}}|$ (not normalized by the gravitational radius). As a consequence, from equations (\ref{eq:AM_p}) and (\ref{eq:energy_p}), one has
\begin{eqnarray}
  && l_{\rm p} \approx - \sqrt{\gamma_{\varphi\varphi}} \hat{u}^t < 0, \\
  && e_{\rm p} \approx (\alpha - \sqrt{\beta^2})\hat{u}^t < 0,
\end{eqnarray}
in the ergosphere, where $\alpha^2 < \beta^2 = \gamma_{\varphi\varphi} \Omega^2$. For $D^2 > B^2$, $\hat{v}^\varphi$ does not approach $-1$, so that $e_{\rm p} > 0$ near the boundary of the ergosphere where $\alpha^2 = \beta^2$. However, $\alpha \to 0$ for $r \to r_{\rm H}$ implies that $e_{\rm p} < 0$ can be realized near the horizon. Here we emphasize that $l_{\rm p}$ and $e_{\rm p}$ are scalars, and thus $l_{\rm p} < 0$ and $e_{\rm p} < 0$ also in the KS coordinates.

For typical AGN jets, $\ell_{\rm gy}$ is expected to be $\sim 10$ orders of magnitude smaller than $GM/c^2$ \citep[cf.][]{toma12}, so that the distance which a particle travels until it achieves the asymptotic azimuthal velocity is tiny compared to the size of the ergosphere. This justifies our calculations of the particle motion in the fixed orthonormal basis with the uniform electromagnetic fields, and the asymptotic velocities can be interpreted as the local velocities of the test particles.

Since the current crossing region is bounded at $r < r_{\rm es}$, the positively charged particles do not cross the last ergospheric field line and will gyrate around this field line. When they emerge out of the ergosphere, they contribute to the current flowing outward along the last ergospheric field line (see Figure~\ref{fig:summary}). The particles outside the ergosphere generally have positive energies.

\subsection{Comparison to the mechanical Penrose process}
\label{sec:penrose}

We argue that BZ process for the ergospheric field lines threading the equatorial plane is similar to the mechanical Penrose process, in which the rotational energy of a BH is extracted as mechanical energy by making the BH absorb negative-energy particles \citep{penrose69,bardeen72}. For simplicity, let us consider the positively and negatively charged particles in the geometrically thin current crossing region as a one-fluid. The energy equation for this fluid in the steady state is written as
\begin{equation}
  \partial_r \sqrt{\gamma} (-\alpha T_{{\rm p},t}^r) = \mathbf{E} \cdot \mathbf{J}_{\rm p} < 0,
  \label{eq:particleEOM}
\end{equation}
where $T_{{\rm p},\mu}^\nu$ is the energy-momentum tensor of the fluid. The boundary condition at $r = r_{\rm es}$ is $T_{{\rm p},t}^r = 0$. Therefore, the solutions of equation (\ref{eq:particleEOM}) should be $F^r \equiv -\alpha T_{{\rm p},t}^r > 0$ in the current crossing region. Then one has $- T_{{\rm p},t}^r = - \rho_{\rm m} U_t U^r > 0$, where $\rho_{\rm m}$ and $U^\mu$ are the comoving mass density and the four-velocity of the fluid, respectively. Since all the particles may have negative energy, it is reasonable to estimate
\begin{equation}
  -U_t < 0, ~~~ U^r < 0.
  \label{eq:penrose}
\end{equation}
This means that the current crossing region generates the inflow of the negative-energy fluid and the outward Poynting flux, which appears to be a similar process to the mechanical Penrose process.

\begin{figure}[t]
\centering\includegraphics[scale=0.5]{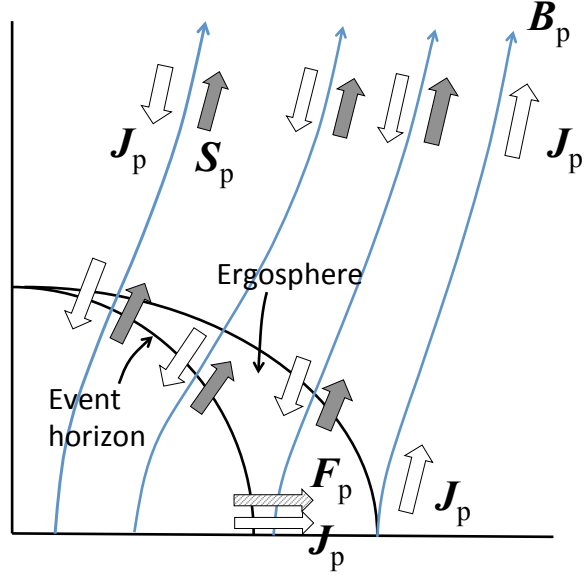}  
\caption{
Schematic picture of the poloidal currents $\mathbf{J}_{\rm p}$ ({\it open arrows}), the poloidal Poynting flux $\mathbf{S}_{\rm p}$ ({\it filled arrows}), and the poloidal particle energy flux $\mathbf{F}_{\rm p}$ near the equatorial plane (i.e. the inflow of the particle negative energies; {\it striped arrow}) in the steady state in the northern hemisphere in the KS coordinates. The BH loses its rotational energy directly by $\mathbf{S}_{\rm p}$ along the field lines threading the horizon (see Sections~\ref{sec:horizon} and \ref{sec:unsteady}) and by $\mathbf{F}_{\rm p}$ near the equatorial plane which is associated with $\mathbf{S}_{\rm p}$ along the field lines threading the equatorial plane in the ergosphere (see Section~\ref{sec:equat}).
}
\label{fig:summary}
\end{figure}

As a result, the BH loses its rotational energy by the poloidal particle energy flux $\mathbf{F}_{\rm p} = -\alpha \rho_{\rm m} U_t \mathbf{U}_{\rm p}$. We summarize our argument in Figure~\ref{fig:summary} (see Section~\ref{sec:horizon} for the field lines threading the horizon).

However, it is too simple to treat the charged particles in the current crossing region as a one-fluid, since the average velocities of the positively and negatively charged particles should be different. Furthermore, Figure~\ref{fig:motcal} is just the result of the test particle calculations. More detailed studies of the plasma particle motions are required to confirm whether the condition of equation (\ref{eq:penrose}) is realistic in the current crossing region.

\subsection{Comparison to MHD numerical simulation results}
\label{sec:MHDnumerical}

MHD numerical simulations treat the energy of particles (while force-free simulations not), so that they can observe the negative particle energy in principle. However, the MHD simulation results of the dilute Kerr BH magnetosphere with cylindrical magnetic field at the far zone in \cite{komissarov05} do not show any negative particle energy in the steady state. This is just due to the disappearance of the ergospheric field lines threading the equatorial plane, although the reason of this disappearance has not been identified. Such behavior is also seen in the three dimensional MHD simulations including the dense accretion flow \citep{tchekho11,mckinney12}\citep[but see][]{punsly15}.

\section{Field lines threading the event horizon}
\label{sec:horizon}

\subsection{Force-free condition is satisfied}
\label{sec:ff_horizon}

In contrast to the equatorial plane where $H_\varphi = 0$ from the symmetry, one generally has $H_\varphi \neq 0$ at the horizon. Thus the above argument on the field lines threading the equatorial plane is not applicable for the field lines threading the horizon. At the horizon, the regularity condition should be satisfied \citep{znajek77,thorne86}:
\begin{equation}
  \hat{B}_\varphi = -\hat{D}_\theta ~~({\rm in~BL~coordinates}).
\label{eq:regularity_BL}  
\end{equation}
For the BZ split-monopole solution \citep{blandford77} as an example, in which $\hat{B}_r \neq 0$, $\hat{B}_\theta \approx 0$, and $\hat{D}_r \approx 0$ (to the zeroth order of $a$), so that one has
\begin{equation}
B^2 - D^2 > 0.
\end{equation}
Therefore the force-free condition is satisfied at the horizon.

We confirm this fact more generally in the KS coordinates. From the calculation shown in Appendix~\ref{app:B2D2}, we obtain
\begin{eqnarray}
  (B^2 - D^2)\alpha^2 &=& -B^\theta B_\theta f(\Omega_{\rm F}, r, \theta) + (B^\varphi B_\varphi + B^r B_r) \frac{\varrho^2 \Delta}{\Sigma} \nonumber \\
  && + 4r \sin^2\theta \left(\Omega_{\rm F} - \Omega \right)B^r B^\varphi
  + \frac{4r^2}{\Sigma} \left[ 1- \left(\frac{\Omega_{\rm F}}{\Omega}\right)^2 \left(1-\frac{\varrho^4}{\Sigma}\right) \right] (B^r)^2,
  \label{eq:B2D2}
\end{eqnarray}
where $\Sigma > \varrho^4$ is generally satisfied (see equation \ref{eq:identity2}). For the BZ split-monopole solution as an example, in which $B^r > 0$ and $B^\varphi < 0$ at the northern hemisphere (see Section~\ref{sec:structure} below), $B^\theta \approx 0$, and $\Omega_{\rm F} \approx \Omega_{\rm H}/2$, one has $B^2 - D^2 > 0$ in the region where $\Omega > \Omega_{\rm F}$. We see that $B^2 - D^2 > 0$ is generally satisfied where $B^\theta$ is weak, $B^r B^\varphi < 0$, and $\Omega > \Omega_{\rm F}$. (Note that for the field lines threading the equatorial plane, $B^\theta$ is the dominant field component near that plane, where $B^2 - D^2 < 0 $ can be realized.)

Therefore, for the field lines threading the horizon, the force-free condition can be satisfied, and then no poloidal current is driven to flow across the field lines in the steady state. No AM or energy is transferred from the particles to the electromagnetic fields. These properties clearly indicate that the flux production mechanism for the field lines threading the horizon is different from that for the field lines threading the equatorial plane.

\begin{figure}
\centering\includegraphics[scale=0.5]{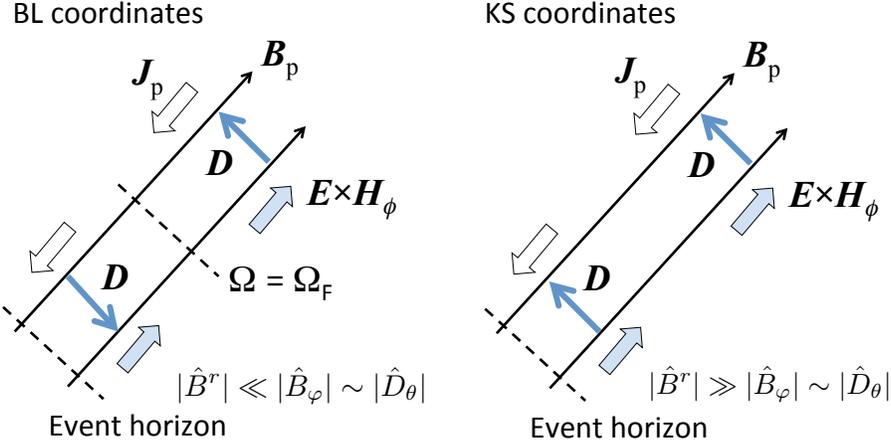}
\caption{
Electromagnetic field structures of the BZ split-monopole solution as measured in the BL ({\it left}) and KS ({\it right}) coordinates.
}
\label{fig:structure}
\end{figure}

\subsection{Electromagnetic structure}
\label{sec:structure}

Here we focus on the electromagnetic structure of the BZ split-monopole solution, and show that some properties are measured differently in the BL and KS coordinates. This analysis is useful for finding the essential physics in BZ process for the field lines threading the horizon, which should be independent of the adopted coordinate systems.

The split-monopole field is given by
\begin{equation}
  B^r = {\rm const.} \times \frac{\sin\theta}{\sqrt{\gamma}}, ~~~ B^\theta \approx 0,
\end{equation}
which satisfies $\nabla \cdot \mathbf{B} = 0$. (Note that $\sin\theta/\sqrt{\gamma} \to 1/r^2$ for $r \to \infty$.) In the BL coordinates, one has
\begin{eqnarray}
 && B_\varphi = \frac{1}{\alpha} H_\varphi, \\
 && D_\theta = \frac{-1}{\alpha} (\Omega_{\rm F} - \Omega) B^r \sqrt{\gamma}.
\end{eqnarray}
As is well known, $D_\theta$ changes its sign at the point where $\Omega = \Omega_{\rm F}$ (see Figure~\ref{fig:structure}, {\it left}). We can see that $B_\varphi$ and $D_\theta$ diverge as $r \to r_{\rm H}$, while $B^r \sqrt{\gamma}$ is finite, and one has
\begin{equation}
  |\hat{B}^r| \ll |\hat{B}_\varphi| \sim |\hat{D}_\theta|
\end{equation}
near the horizon.

On the other hand, in the KS coordinates, one has
\begin{eqnarray}
  && B^\varphi = \frac{\alpha H_\varphi -  B^r \sin^2\theta (2r\Omega_{\rm F} - a)}{\Delta \sin^2\theta}, \label{eq:Bphi_KS} \\  
  && D_\theta = \frac{-1}{\alpha} (\Omega_{\rm F} B^r - \beta^r B^\varphi) \sqrt{\gamma}.
\end{eqnarray}
Equation (\ref{eq:Bphi_KS}) is derived by rewriting $H_\varphi$ in equation (\ref{eq:relation_H}) with equation (\ref{eq:relation_E}) and (\ref{eq:identity3}) (K04). The regularity condition at the horizon ($\Delta = 0$) for the steady flow to pass with no diverging physical quantities is given by
\begin{equation}
  \alpha H_\varphi =  B^r \sin^2\theta (2r\Omega_{\rm F} - a),
  \label{eq:regularity_KS}
\end{equation}
which is equivalent to equation (\ref{eq:regularity_BL}). We calculate $B^\varphi$ and $D_\theta$ from $r = r_{\rm H}$ towards infinity for small values of $a$ and $\Omega_{\rm F} = \Omega_{\rm H}/2$, and find that $B^\varphi < 0$, that $D_\theta$ does not change its sign (see Figure~\ref{fig:structure}, {\it right}), and that
\begin{equation}
  |\hat{B}^r| \gg |\hat{B}_\varphi| \sim |\hat{D}_\theta|
\label{eq:KSfield}
\end{equation}
near the horizon. That is, the $\mathbf{D}$ field in the KS coordinates is not only so weak that it cannot drive the cross-field current but also it does not change its direction, i.e. $\mathbf{D} \cdot \mathbf{E} > 0$ in the whole region. This situation is in stark contrast to the field lines threading the equatorial plane, for which the cross-field current is driven by the strong $\mathbf{D}$ field with $\mathbf{D} \cdot \mathbf{E} < 0$.

In the BL coordinates the point where $\Omega = \Omega_{\rm F}$ and $\mathbf{D} = 0$ appears special, and it was considered as a key in some previous analytical discussions \citep[e.g.][K09]{okamoto06}. However, the electromagnetic quantities are clearly continuous or seamless in the KS coordinates, as shown in Figure~\ref{fig:structure}.

Below we generically consider the cases in which the force-free condition is satisfied along the field lines threading the horizon (see Section~\ref{sec:ff_horizon}). In those cases, an essential point is that the outward AM and Poynting fluxes, $\mathbf{L}_{\rm p} = -H_\varphi \mathbf{B}_{\rm p}/4\pi$ and $\mathbf{S}_{\rm p} = -\Omega_{\rm F} H _\varphi \mathbf{B}_{\rm p}/4\pi$, are seamless along each field line from the event horizon to infinity in the steady state (from equation \ref{eq:forcefree}), with no transfer of AM and energy from the particles. This situation is described in Figure~\ref{fig:summary}.

\subsection{The issue}
\label{sec:problem}

Now we discuss how $\mathbf{L}_{\rm p}$ and $\mathbf{S}_{\rm p}$ are created along the field lines threading the horizon. Blandford \& Znajek \cite{blandford77} show that $\Omega_{\rm F}$ and $H_\varphi$ in the steady state are determined {\it mathematically} from equations (\ref{eq:omegaF}) and (\ref{eq:forcefree}) with the conditions at the horizon and at infinity (see also K04). This mathematics and the seamless property shown above may lead to an incorrect consideration that the fluxes are created at the horizon. The conditions at the horizon and at infinity are not boundary conditions but regularity conditions \citep[][K04]{beskin00}, as stated in Section~\ref{sec:intro}. The place where the fluxes are created must not be the horizon, but outside the horizon.

We note that the non-zero outward AM and Poynting fluxes at the horizon in the KS coordinates do not violate causality, because the steady fluxes carry no information. It should be also noted that the steady Poynting flux $\mathbf{S}_{\rm p} = -\Omega_{\rm F} H_\varphi \mathbf{B}_{\rm p}/4\pi$ is not a product of a certain energy density and its advection speed like steady particle energy flux $\mathbf{F}_{\rm p} = (-\alpha \rho_m U_t) \mathbf{U}_{\rm p}$ (see Section~\ref{sec:negative_EM} for a related discussion). The Poynting flux is just a result of the currents flowing in the plasma with the potential differences.

Consequently, the issue on the field lines threading the horizon is well defined as ``How is the steady current structure causally built?" We consider that this issue may not be resolved by investigating only the steady-state structure. The phenomena at the horizon should be a result from those having occurred outside the horizon in the {\it prior} times $t$. In Section~\ref{sec:unsteady}, we address this issue by discussing a time-dependent state evolving towards the steady state.

\subsection{Negative electromagnetic energy?}
\label{sec:negative_EM}

Lasota et al. \cite{lasota14} and Koide \& Baba \cite{koide14} argue that the outward Poynting flux is mediated by `inflow of the negative electromagnetic energy' (see also K09). Although this interpretation analogous to the mechanical Penrose process looks attractive for causal production of the Poynting flux, it is difficult to consider {\it the flow of the steady field} (rather than waves). Furthermore, we find that the sign of the electromagnetic energy density depends on the coordinates. 

In the BL coordinates, the electromagnetic AM and energy densities can be written down by (K09)
\begin{eqnarray}
  && l = \frac{1}{4\pi\alpha} \gamma_{\varphi\varphi} (\Omega_{\rm F} - \Omega) (B^\theta B_\theta + B^r B_r), \\ 
  && e = \frac{1}{8\pi\alpha}\left[\alpha^2 B^2 + \gamma_{\varphi\varphi}(\Omega_{\rm F}^2 - \Omega^2)(B^\theta B_\theta + B^r B_r) \right].
\end{eqnarray}
Thus $l$ and $e$ is negative (and diverges) near the horizon when $\Omega_{\rm F} < \Omega_{\rm H}$. This condition is satisfied in the BZ split-monopole solution.

On the other hand, in the KS coordinates, the calculations shown in Appendix~\ref{app:B2D2} lead to
\begin{eqnarray}
  4\pi\alpha l &=& \frac{\Sigma \sin^2\theta}{\varrho^2} (\Omega_{\rm F} - \Omega) B^\theta B_\theta
  -2r \sin^2\theta B^rB^\varphi
  + \Omega_{\rm F} (\varrho^2 + 2r)\sin^2\theta (B^r)^2
   \label{eq:l_KS}\\
  8\pi\alpha e &=& \left[\frac{\Sigma \sin^2\theta}{\varrho^2} (\Omega_{\rm F} + \Omega)(\Omega_{\rm F} - \Omega) + \frac{\varrho^2 \Delta}{\Sigma} \right] B^\theta B_\theta
  + \Delta \sin^2\theta (B^\varphi)^2 \nonumber \\
  && - 2a\sin^2\theta B^r B^\varphi
  + \left[1+\Omega_{\rm F}^2 (\varrho^2 + 2r)\sin^2\theta \right] (B^r)^2.
  \label{eq:e_KS}
\end{eqnarray}
In the BZ split-monopole solution as an example, in which $B^\theta \approx 0$ and $B^\varphi < 0$ in the northern hemisphere, one has
\begin{equation}
  l > 0, ~~~ e > 0.
\end{equation}
This condition is generally valid when $B^\theta$ is weak and $B^r B^\varphi < 0$.

Note that
\begin{equation}
  l = \alpha T^t_\varphi = -T^\mu_\nu n_\mu \chi^\nu, ~~~
  e = -\alpha T^t_t = T^\mu_\nu n_\mu \xi^\nu
  \label{eq:le_n}
\end{equation}
depend on the coordinates, while $T_{t\varphi} = T_{\mu\nu} \xi^\mu \chi^\nu$ and $T_{tt} = T_{\mu\nu} \xi^\mu \xi^\nu$ are scalars. The concept of the negative electromagnetic energy density depends on the coordinates, and thus it is not physically essential.\footnote{Lasota et al. \cite{lasota14} argue that the electromagnetic energy density calculated in the KS coordinates is negative near the horizon, but they define the electromagnetic energy density as $T_{\mu\nu} l^{\mu} \xi^{\nu}$ where $l^\mu = \alpha n^\mu$ and $n^\mu$ is the four-velocity of the BL FIDO.}

\section{Process towards the Steady State}
\label{sec:unsteady}

As stated in Section~\ref{sec:problem}, we address the issue how the steady poloidal current structure is built causally, by discussing a time-dependent state evolving towards the steady state.

In the steady state, the plasma has the inner and outer light surfaces (see Section~\ref{sec:basic_motion}). The particles flow in across the inner light surface and flow out across the outer light surface. Therefore, new particles have to keep injected between the two light surfaces, as discussed in many literatures \citep[e.g.][]{takahashi90,beskin00,mckinney06,broderick15}. In this paper we have assumed that the plasma particles keep injected from outside the magnetosphere through electron-positron pair creation by collisions of two photons \citep{mckinney06,levinson11,moscibrodzka11} and/or diffusion of high-energy hadrons \citep{toma12}\footnote{In the geometrically thick accretion disk the particles can be non-thermally accelerated and diffused out of the disk. The amount of those high-energy hadrons does not appear to be sufficient for the total mass loading of AGN jets which provides the observationally inferred Lorentz factor $\Gamma \sim 10-100$, but sufficient for satisfying $\mathbf{D} \cdot \mathbf{B} = 0$ \citep{kimura14,kimura15}.}, and that those particles maintain $\mathbf{D} \cdot \mathbf{B} = 0$ and carry the currents.

\begin{figure}[t]
\centering\includegraphics[scale=0.5]{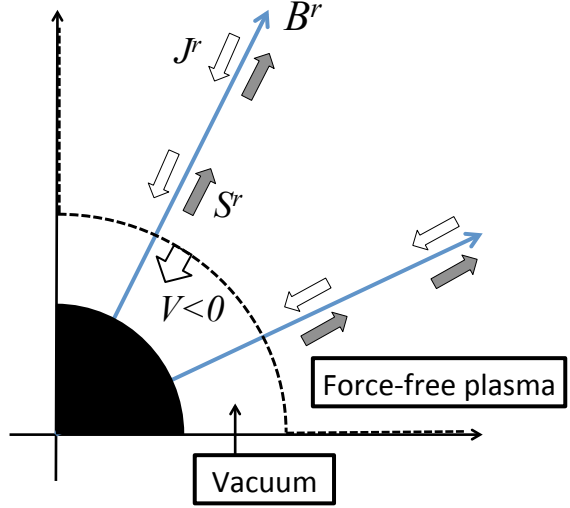}  
\caption{
Schematic picture of a time-dependent process evolving towards the steady state. The plasma particles keep injected between the inner and outer light surfaces, and the vacuum is being filled with those plasma. This picture focuses on the inflow. The inner boundary of the force-free region propagates towards the event horizon, producing the steady poloidal current structure and the outward AM and Poynting fluxes.
}
\label{fig:unsteady}
\end{figure}

Now let us first consider a {\it vacuum} in Kerr space-time, and then begin the continuous injection of force-free plasma particles between the two light surfaces as a gedankenexperiment. The inflow (outflow) will fill the vacuum near the horizon (at infinity). Simultaneously we will see a process building the poloidal current structure. Hereafter we will call the (inflow + outflow) region filled with the force-free plasma `force-free region'. Figure~\ref{fig:unsteady} is a schematic picture of this process focusing on the inflow.

\begin{figure}[t]
\centering\includegraphics[scale=0.5]{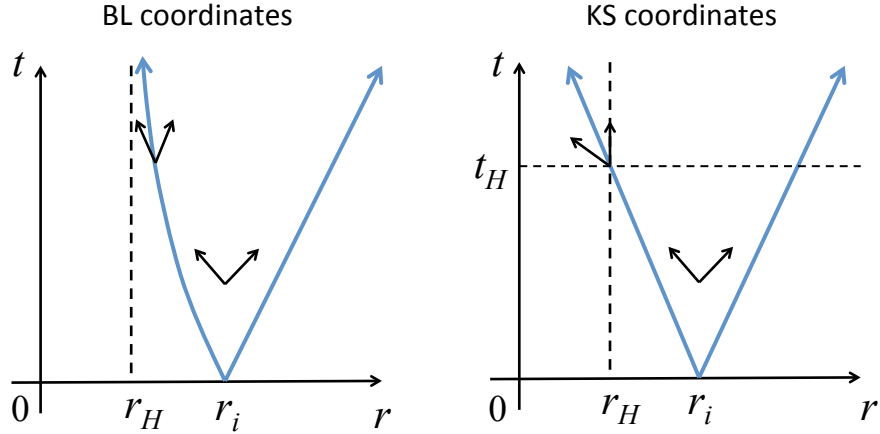}
\caption{
Space-time diagram of the inner and outer boundaries of the force-free region in the BL and KS coordinates. In each diagram the left and right long arrows correspond to the motions of the inner and outer boundaries, respectively, while the small arrows to the propagation of light.
}
\label{fig:spacetime}
\end{figure}

We show the space-time diagrams of the inner and outer boundaries of the force-free region in the BL and KS coordinates in Figure~\ref{fig:spacetime}, in which the radial light signals are represented by the small arrows. The outflow continues to propagate into the vacuum, i.e. the radius of the outer boundary $r \to \infty$ for $t \to \infty$. In the BL coordinates, the inflow also continues to propagate towards the horizon, $r \to r_{\rm H}$ for $t \to \infty$. In the KS coordinates, the inflow can pass the horizon in a finite time of $t = t_{\rm H}$. In both of the coordinates, when the inner boundary approaches the horizon, the outward signal from it becomes slower and slower and it can hardly affect the force-free region. This will lead to the steady state.\footnote{In some MHD simulations, a static plasma (not a vacuum) is initially given and then a central star starts rotating \citep{bogovalov99} or a BH starts rotating \citep{komissarov04b}. They show that a switching-on wave propagates outward and that the outflow region settles down to the steady state after it passes the outer fast magnetosonic point \citep{beskin10}.}

Although such a time-dependent state should be analyzed numerically, we use a toy model to qualitatively illustrate the process of building the poloidal current structure. This model assumes that (1) $\mathbf{B}_{\rm p}$ is fixed to be split-monopole
\begin{equation}
  \partial_r (\sqrt{\gamma} B^r) = 0, ~~~ B^\theta = 0
\end{equation}
in the whole region, and that (2) the Kerr BH magnetosphere is separated into the force-free region and the vacuum by geometrically thin boundaries moving radially. For further simplicity, (3) we assume that the force-free region and the vacuum have their steady-state structures, but the values of the physical quantities, particularly $\Omega_{\rm F}$ and $H_\varphi$, keep updated as determined by the varying conditions of the inner and outer boundaries.

Some of these assumptions would be violated in realistic experiments. Nevertheless we consider that our toy model is useful to suggest the key points for resolving the issue on the causality in the coordinate basis (Section~\ref{sec:flux_production}), which also allows us to understand how the steady state is maintained (Section~\ref{sec:discussion}).

\subsection{Analysis in the BL coordinates}

\subsubsection{The force-free and vacuum regions}

The electromagnetic quantities in the force-free region are given as follows. The condition $\mathbf{D} \cdot \mathbf{B} = 0$ and $\nabla \times \mathbf{E} = 0$ lead to
\begin{equation}
  E^{\rm ff}_\varphi = E^{\rm ff}_r = 0, ~~~ E^{\rm ff}_\theta = - \sqrt{\gamma} \Omega_{\rm F} B^r,
\end{equation}
where
\begin{equation}
  \partial_r \Omega_{\rm F} = 0.
\end{equation}
Hereafter we will put the subscript and superscript `ff' on the quantities in the force-free region. Equations (\ref{eq:relation_E}) and (\ref{eq:relation_H}) give us
\begin{equation}
  D^{\rm ff}_\varphi = D^{\rm ff}_r = 0, ~~~ D^{\rm ff}_\theta = \frac{\sqrt{\gamma}}{\alpha} (\Omega - \Omega_{\rm F}) B^r,
\end{equation}
\begin{equation}
  H^{\rm ff}_\varphi = \alpha B^{\rm ff}_\varphi, ~~~ H^{\rm ff}_r = \alpha B_r - \sqrt{\gamma} \Omega D_{\rm ff}^\theta, ~~~ H^{\rm ff}_\theta = 0.
\end{equation}
Equation $\nabla \times \mathbf{H} = 4\pi \mathbf{J}$ and the force-free condition lead to
\begin{equation}
  \partial_r H^{\rm ff}_\varphi = -4\pi \sqrt{\gamma} J_{\rm ff}^\theta = 0,
\end{equation}
\begin{equation}
  \partial_\theta H^{\rm ff}_\varphi = 4\pi \sqrt{\gamma} J^r_{\rm ff},
  \label{eq:ampere_ff_BL}
\end{equation}
These two equations imply that $\partial_r (\sqrt{\gamma} J_{\rm ff}^r) = 0$. We focus on the northern hemisphere, where $J_{\rm ff}^r < 0$ and $H^{\rm ff}_\varphi < 0$. The current flowing outward $J_{\rm ff}^r > 0$, which prevents the BH from charging up, is assumed to be concentrated on the equatorial plane. The poloidal AM and Poynting fluxes are
\begin{equation}
  L_{\rm ff}^r = \frac{-H^{\rm ff}_\varphi}{4\pi} B^r, ~~~ S_{\rm ff}^r = \Omega_{\rm F} \frac{-H^{\rm ff}_\varphi}{4\pi} B^r,
\end{equation}
which satisfy $\partial_r (\sqrt{\gamma} L_{\rm ff}^r) = 0$ and $\partial_r (\sqrt{\gamma} S_{\rm ff}^r) = 0$.

In the vacuum region, one has $\rho = \mathbf{J} = 0$. Equations $\nabla \times \mathbf{E} = 0$ and $\nabla \times \mathbf{H} = 0$ lead to
\begin{equation}
  E^{\rm vac}_\varphi = 0, ~~~ H^{\rm vac}_\varphi = B^{\rm vac}_\varphi = 0,
\end{equation}
which indicates
\begin{equation}
  L^r_{\rm vac} = S^r_{\rm vac} = 0.
\end{equation}
Hereafter we will put the subscript and superscript `vac' on the quantities in the vacuum region.

\subsubsection{The inner boundary of the force-free region}
\label{sec:innerb}

Let us focus on the inner boundary of the force-free (inflow) region, and derive the conditions on the boundary, i.e. the junction conditions between the force-free and vacuum regions. The similar analysis can be done for the outer boundary. For equation
\begin{equation}
  -\partial_t D^r + \frac{1}{\sqrt{\gamma}} \partial_\theta H_\varphi = 4\pi J^r,
  \label{eq:ampere_BL}
\end{equation}
 we substitute
\begin{eqnarray}
  D^r &=& D_{\rm vac}^r H(-R), \\
  H_\varphi &=& H^{\rm ff}_\varphi H(R), \label{eq:Hphi_BL} \\
  J^r &=& J_{\rm ff}^r H(R) + \eta^r \delta(R) \label{eq:Jr_BL},
\end{eqnarray}
where $H(R)$ and $\delta(R)$ are the Heaviside step function and the Dirac delta function, respectively, and
\begin{equation}
  R = r - r_i - \int^t_0 V dt,
\end{equation}
where $r_i$ and $V$ are the initial radius and the velocity of the boundary. The location of the boundary is represented by $R=0$. We have introduced $\eta^r$ in equation (\ref{eq:Jr_BL}), i.e. possible contribution to $J^r$ from moving surface charges at the boundary. The assumption (3) stated in the first part of this section implies that the timescale for the quantities in the force-free and vacuum regions becoming adjusted for steady-state structure is much smaller than the timescale of the boundary propagation. We focus on the latter timescale, considering that only $R = R(t)$ depends on $t$ in equation (\ref{eq:ampere_BL}). Then we have
\begin{equation}
  -D_{\rm vac}^r V \delta(R) + \frac{1}{\sqrt{\gamma}} (\partial_\theta H_\varphi^{\rm ff}) H(R) = 4\pi J_{\rm ff}^r H(R) + 4\pi \eta^r \delta(R).
\end{equation}
Taking account of equation (\ref{eq:ampere_ff_BL}), we obtain
\begin{equation}
  \eta^r = \left. \frac{-D_{\rm vac}^r}{4\pi} \right|_{R=0} V,
  \label{eq:etar_BL}
\end{equation}
which implies that the surface charge density on the boundary $\sigma = -D_{\rm vac}^r|_{R=0}/4\pi$. This can be confirmed by integrating $\nabla \cdot \mathbf{D} = 4\pi \rho$ over the infinitesimally thin (in the $r$ direction) region enclosing the small area on the boundary and taking account of $D_{\rm ff}^r = 0$.

For equation
\begin{equation}
  -\partial_t D^\theta - \frac{1}{\sqrt{\gamma}} \partial_r H_\varphi = 4\pi J^\theta,
  \label{eq:ampere2_BL}
\end{equation}
we substitute
\begin{eqnarray}
  D^\theta &=& D_{\rm vac}^\theta H(-R) + D_{\rm ff}^\theta H(R), \\
  J^\theta &=& \eta^\theta \delta(R)
  \label{eq:Jtheta_BL},
\end{eqnarray}
and equation (\ref{eq:Hphi_BL}). We have introduced $\eta^\theta$, possible contribution to $J^\theta$ from the surface current flowing on the boundary. Then we have
\begin{equation}
  -D_{\rm vac}^\theta V \delta(R) + D_{\rm ff}^\theta V \delta(R) - \frac{1}{\sqrt{\gamma}} H_\varphi^{\rm ff} \delta(R) = 4\pi \eta^\theta \delta(R),
\end{equation}
which leads to
\begin{equation}
  V = \frac{1}{\sqrt{\gamma}} \left. \frac{H_\varphi^{\rm ff} + 4\pi \sqrt{\gamma} \eta^\theta}{D_{\rm ff}^\theta - D_{\rm vac}^\theta} \right|_{R=0}.
  \label{eq:V1_BL}
\end{equation}

The last one of Maxwell equations nontrivial for the present problem is
\begin{equation}
  \partial_t B^\varphi + \frac{1}{\sqrt{\gamma}} (\partial_r E_\theta - \partial_\theta E_r) = 0,
  \label{eq:faraday_BL}
\end{equation}
for which we substitute
\begin{eqnarray}
  B^\varphi &=& B_{\rm ff}^\varphi H(R), \\
  E_\theta &=& E_\theta^{\rm vac} H(-R) + E_\theta^{\rm ff} H(R), \\
  E_r &=& E_r^{\rm vac} H(-R).
\end{eqnarray}
Then we have
\begin{eqnarray}
  -B_{\rm ff}^\varphi V \delta(R) + \frac{1}{\sqrt{\gamma}} \left[-E_\theta^{\rm vac} \delta(R) + E_\theta^{\rm ff} \delta(R)
  - (\partial_\theta E_r^{\rm vac})H(-R) \right] = 0.
  \label{eq:faraday_BL2}
\end{eqnarray}
Integrating equation (\ref{eq:faraday_BL2}) over $-\epsilon < R < \epsilon$ and take a limit of $\epsilon \to 0$, the last term vanishes, and we obtain
\begin{eqnarray}
  V &=& \frac{1}{\sqrt{\gamma}} \left. \frac{E_\theta^{\rm ff} - E_\theta^{\rm vac}}{B^\varphi_{\rm ff}} \right|_{R=0}, \nonumber \\
  &=&  \frac{\alpha}{\sqrt{\gamma}} \left. \frac{D_\theta^{\rm ff} - D_\theta^{\rm vac}}{B^\varphi_{\rm ff}} \right|_{R=0},
  \label{eq:V2_BL}
\end{eqnarray}
where we have used equation (\ref{eq:relation_E}) for the last equality. Eliminating $D_{\rm ff}^\theta - D_{\rm vac}^\theta$ from equations (\ref{eq:V1_BL}) and (\ref{eq:V2_BL}) leads to
\begin{equation}
  V = \frac{\pm\alpha}{\sqrt{\gamma_{rr}}} \sqrt{1+\frac{4\pi \sqrt{\gamma} \eta^\theta}{H_\varphi^{\rm ff}}}.
\end{equation}
Here we take the minus sign, since we have assumed that the inner boundary keeps moving inward. In Section~\ref{sec:consistency}, we will confirm that this assumption is consistent with the electromagnetic structure which we found.

Let us consider the case of $\eta^\theta = 0$. Then we have
\begin{equation}
  V = \frac{-\alpha}{\sqrt{\gamma_{rr}}},
  \label{eq:Vmax_BL}
\end{equation}
and 
\begin{eqnarray}
  H_\varphi^{\rm ff} &=& -\alpha \left. \sqrt{\frac{\gamma_{\varphi\varphi}}{\gamma_{\theta\theta}}} (D_\theta^{\rm ff} - D_\theta^{\rm vac}) \right|_{R=0} \nonumber \\
  &=& -\left. \sqrt{\frac{\gamma_{\varphi\varphi}}{\gamma_{\theta\theta}}} \left[(\Omega - \Omega_{\rm F}) \sqrt{\gamma} B^r - \alpha D_\theta^{\rm vac} \right] \right|_{R=0}.
  \label{eq:znajek_BL}
\end{eqnarray}
Substituting $dr = Vdt$ for equation (\ref{eq:metric}), we find
\begin{equation}
ds^2 = \gamma_{\varphi\varphi}(d\varphi - \Omega dt)^2 + \gamma_{\theta\theta}d\theta^2 \ge 0,
\end{equation}
which has to be $ds^2 = 0$. This means that the four-velocity of the boundary is null. In reality, however, the particles at the boundary cannot propagate with this speed, and thus one can conclude
\begin{equation}
  \eta^\theta > 0,
  \label{eq:CFcurrent_BL}
\end{equation}
i.e., the cross-field current must flow on the boundary. Note that equation (\ref{eq:znajek_BL}) with $\alpha D^{\rm vac}_\theta \to 0$ becomes equivalent to the regularity condition at the horizon (equation \ref{eq:regularity_BL}).

\subsubsection{Consistency check}
\label{sec:consistency}

In our toy model of the time-dependent state, we have not taken into account equations of the particle motions, using the force-free approximation for the force-free region, but we have assumed that the inner boundary keeps moving inward, i.e. $V<0$. Here we examine the direction of the Lorentz force exerted on the particles at the boundary, and confirm that it is consistent with the assumption of $V < 0$. It is reasonable that the force-free approximation is not applicable for the boundary between the force-free and vacuum regions, and indeed we have seen that the cross-field current flows there, $\eta^\theta > 0$.

The particle number density $n_{\rm ff}$ of the force-free region is high enough to screen the electric field along the $\mathbf{B}$ field lines, i.e. $D^r_{\rm ff} = 0$. We may even assume that $n_{\rm ff} \gg \rho_{\rm ff}/e$, where $\rho_{\rm ff}$ is the charge density of the force-free region, and then the distribution of $n_{\rm ff}$ is not directly related to that of $\rho_{\rm ff}$. On the other hand, $n$ approaches zero at the boundary towards the vacuum region, where $n \gg \rho/e$ is not valid, and non-zero surface charge density $\sigma$ just implies non-zero surface mass density $\sigma_{\rm m}$. Thus we can write the equation of the particle motions in the $r$ direction as $\nabla_\nu [\sigma_{\rm m} U^r U^\nu \delta(R) + \rho_{\rm mff} U^r_{\rm ff} U^\nu_{\rm ff} H(R)] = F^r_{~~\nu} I^\nu$ and the continuity equation as $\nabla_\nu [\sigma_{\rm m} U^\nu \delta(R) + \rho_{\rm mff} U^\nu_{\rm ff} H(R)] = 0$, where $\rho_{\rm mff}$ is the mass density of the force-free region. We combine these two equations, use $U_{\rm ff}^\nu \partial_\nu H(R) = U^t_{\rm ff}(V^r_{\rm ff} - V)\delta(R)$, and integrate the equation over $R$ (i.e., keep the components including $\delta(R)$ as done in Section~\ref{sec:innerb}) to have
\begin{equation}
  \sigma_{\rm m} U^\nu \nabla_\nu U^r + \rho_{\rm mff} U^t_{\rm ff} (V_{\rm ff}^r - V) (U^r_{\rm ff} - U^r) = \sigma D^r|_{R=0} + \frac{\gamma_{\theta\theta}}{\alpha \sqrt{\gamma}} \eta^\theta B_\varphi|_{R=0}.
  \label{eq:EOM_boundary}
\end{equation}
At the boundary $D^r \neq 0$ and $n \sim \rho/e$, and then the Lorentz force will be much stronger than the gravitational and inertial forces. We neglect the latter forces as in Section~\ref{sec:equa_eom}, so that the first term in the left-hand side of equation (\ref{eq:EOM_boundary}) can be rewritten as $\sigma_{\rm m} U^\nu \partial_\nu U^r$.  In equation (\ref{eq:EOM_boundary}), $D^r|_{R=0}$ should have a value between $D^r_{\rm ff} = 0$ and $D^r_{\rm vac} = -4\pi \sigma$, and $B_{\varphi}|_{R=0}$ between $H_\varphi^{\rm ff}/\alpha < 0$ and $H_\varphi^{\rm vac}/\alpha = 0$. We also found $\eta^\theta > 0$. These mean that the right-hand side of equation (\ref{eq:EOM_boundary}) is negative, i.e., the Lorentz force exerted on the boundary is in the direction of $-r$. 

The second term in the left-hand side of equation (\ref{eq:EOM_boundary}) represents momentum change of the boundary layer due to its mass exhange with the force-free region, and this term is zero when $V = V_{\rm ff}^r$. In the other case, we have $V > V_{\rm ff}^r$ since the boundary and the force-free region do not separate. In our toy model, the particles are continuously injected between the two light surfaces, and the particles flow outward across the outer light surface and flow inward across the inner light surface. For the inflowing force-free region, the continuity equation $\nabla_\nu (\rho_{\rm mff} U^\nu_{\rm ff}) = 0$ and its assumed steady axisymmetric structure mean $U^r_{\rm ff} < 0$. If $U^r > U_{\rm ff}^r$, the acceleration $\sigma_{\rm m} U^\nu \partial_\nu U^r$ is negative, and we continue to have $U^r < 0$ and then $V<0$. The acceleration could be positive when $U^r < U^r_{\rm ff} < 0$, but this case means $V<0$. The acceleration cannot be positive while $V>0$. Therefore, the inner boundary keeps $V<0$.

We can also confirm $V>0$ for the outer boundary. Equations (\ref{eq:etar_BL}), (\ref{eq:V1_BL}), and (\ref{eq:V2_BL}) are valid with changes $\eta^r \to -\eta^r$ and $\eta^\theta \to -\eta^\theta$, and thus one obtains the conditions $D^r_{\rm vac}|_{R=0} = 4\pi \sigma$ and $\eta^\theta < 0$. These indicate $F^r_{~~\nu} I^\nu > 0$. Equation (\ref{eq:EOM_boundary}) is valid with change $V^r_{\rm ff} - V \to V - V^r_{\rm ff}$. The same argument as for the inner boundary leads to the conclusion $V > 0$.

In order to check the consistency of our model more rigorously, fully time-dependent numerical calculations without the assumptions that we set are required, but they are beyond the scope of this paper.

\subsubsection{Causal production of the AM and Poynting fluxes}
\label{sec:flux_production}

Since $V < 0$ for the inner boundary and $B^\varphi_{\rm ff} = B_\varphi^{\rm ff}/\gamma_{\varphi\varphi} = H_\varphi^{\rm ff}/\alpha \gamma_{\varphi\varphi} < 0$, equation (\ref{eq:V2_BL}) means
\begin{equation}
  D_\theta^{\rm ff}|_{R=0} > D_\theta^{\rm vac}|_{R=0}.
  \label{eq:Dshift_BL}
\end{equation}
The electromagnetic AM density is given as $l = -D^\theta B^r \sqrt{\gamma}/4\pi$. Then equation (\ref{eq:Dshift_BL}) indicates
\begin{equation}
  l_{\rm ff}|_{R=0} < l_{\rm vac}|_{R=0}.
  \label{eq:AMineq_BL}
\end{equation}
That is, the inner boundary of the force-free region converts the vacuum with larger AM density into the force-free plasma with smaller AM density. Now equation (\ref{eq:AM}) can be written as
\begin{equation}
  B^r \partial_r \left(\frac{-H_\varphi}{4\pi}\right) = -\partial_t l + \sqrt{\gamma} J^\theta B^r.
  \label{eq:AM_BL}
\end{equation}
Substituting 
\begin{equation}
  l = l_{\rm vac} H(-R) + l_{\rm ff} H(R)
\end{equation}
and equations (\ref{eq:Hphi_BL}) and (\ref{eq:Jtheta_BL}) for equation (\ref{eq:AM_BL}), we obtain
\begin{equation}
  L_{\rm ff}^r = \left[ V(l_{\rm ff} - l_{\rm vac}) + \sqrt{\gamma} \eta^\theta B^r \right]_{R=0}.
  \label{eq:AMconv_BL}
\end{equation}
Taking account of equations (\ref{eq:CFcurrent_BL}) and (\ref{eq:AMineq_BL}), we find that the electromagnetic AM flux in the force-free region is produced by the conversion of the electromagnetic AM density from the vacuum to the force-free plasma through the boundary and the torque of the cross-field current at the boundary.

Equation (\ref{eq:AMconv_BL}) can also be derived from equation (\ref{eq:V1_BL}). These equations also mean that $H_\varphi^{\rm ff}$ is produced by the displacement current $\sqrt{\gamma} V (D_{\rm ff}^\theta - D_{\rm vac}^\theta)$ and the cross-field current $-4\pi \sqrt{\gamma} \eta^\theta$. None of these two contributions appears in the steady state (see Sections~\ref{sec:ff_horizon} and \ref{sec:structure}).

Equation (\ref{eq:energy}) can be reduced to
\begin{equation}
  B^r \partial_r \left(\Omega_{\rm F} \frac{-H_\varphi}{4\pi}\right) = -\partial_t e - E_r J^r - E_\theta J^\theta.
  \label{eq:energy_BL}
\end{equation}
Substituting
\begin{equation}
  e = e_{\rm vac} H(-R) + e_{\rm ff} H(R)
\end{equation}
and equations (\ref{eq:Hphi_BL}), (\ref{eq:Jr_BL}), and (\ref{eq:Jtheta_BL}) for equation (\ref{eq:energy_BL}), we obtain
\begin{equation}
  S_{\rm ff}^r = \left[ V(e_{\rm ff} - e_{\rm vac}) - E_r \eta^r - E_\theta \eta^\theta \right]_{R=0}.
  \label{eq:energyconv_BL}
\end{equation}
By using the expressions
\begin{eqnarray}
  e_{\rm ff} &=& \frac{1}{8\pi} (E^{\rm ff}_\theta D_{\rm ff}^\theta + B_{\rm ff}^\varphi H_\varphi^{\rm ff} + B^r H_r), \\
  e_{\rm vac} &=& \frac{1}{8\pi} (E^{\rm vac}_r D_{\rm vac}^r + E^{\rm vac}_\theta D_{\rm vac}^\theta + B^r H_r),
\end{eqnarray}
we find that
\begin{equation}
  E_r|_{R=0} = \left. \frac{E_r^{\rm ff} + E_r^{\rm vac}}{2} \right|_{R=0}, ~~~
  E_\theta|_{R=0} = \left. \frac{E_\theta^{\rm ff} + E_\theta^{\rm vac}}{2} \right|_{R=0}
  \label{eq:E0_BL}
\end{equation}
satisfy equation (\ref{eq:energyconv_BL}). In equation (\ref{eq:energyconv_BL}), the term $-E_r \eta^r|_{R=0} = \alpha D^r_{\rm vac} D_r^{\rm vac} V/8\pi < 0$. One can see that the Poynting flux in the force-free region is produced by the electromagnetic energy conversion $V(e_{\rm ff} - e_{\rm vac})|_{R=0}$ and the work of the cross-field current $-E_\theta \eta^\theta|_{R=0}$.

\subsection{Analysis in the KS coordinates}

We can obtain the same conclusions as above in the KS coordinates, where the calculations are complicated compared to those in the BL coordinates due to $\gamma_{r\varphi} \neq 0$. Equations having different shapes from those in the BL coordinates are
\begin{eqnarray}
  && D_\theta^{\rm ff} = \frac{1}{\alpha} (-\sqrt{\gamma} \Omega_{\rm F} B^r + \sqrt{\gamma} \beta^r B^\varphi), \\
  && H_\varphi^{\rm ff} = \alpha B_\varphi^{\rm ff} - \sqrt{\gamma} \beta^r D^\theta_{\rm ff}, ~~~ H_r^{\rm ff} = \alpha B_r, \label{eq:Hff_KS} 
\end{eqnarray}
for the force-free region, and
\begin{equation}
  H_\varphi^{\rm vac} = \alpha B_\varphi^{\rm vac} - \sqrt{\gamma} \beta^r D^\theta_{\rm vac} = 0,
  \label{eq:Hvac_KS}
\end{equation}
for the vacuum region. From equations (\ref{eq:ampere_BL}), (\ref{eq:ampere2_BL}), and (\ref{eq:faraday_BL}), we obtain
\begin{equation}
  \eta^r = \left. \frac{-D_{\rm vac}^r}{4\pi} \right|_{R=0} V,
\end{equation}
\begin{eqnarray}
  && V = \left. \frac{1}{\sqrt{\gamma}} \frac{H_\varphi^{\rm ff} + 4\pi \sqrt{\gamma} \eta^\theta}{D_{\rm ff}^\theta - D_{\rm vac}^\theta} \right|_{R=0} \label{eq:V1_KS} \\
  && ~~~ = \left[ \frac{1}{\sqrt{\gamma}} \frac{\alpha (B_\varphi^{\rm ff} - B_\varphi^{\rm vac}) + 4\pi \sqrt{\gamma} \eta^\theta}{D_{\rm ff}^\theta - D_{\rm vac}^\theta} - \beta^r \right]_{R=0}, \label{eq:V1_KS2}
\end{eqnarray}
\begin{eqnarray}
  && V = \left. \frac{1}{\sqrt{\gamma}} \frac{E_\theta^{\rm ff} - E_\theta^{\rm vac}}{B_{\rm ff}^\varphi - B_{\rm vac}^\varphi} \right|_{R=0} \label{eq:V2_KS} \\
  && ~~~ = \left(\frac{\alpha}{\sqrt{\gamma}} \frac{D_\theta^{\rm ff} - D_\theta^{\rm vac}}{B_{\rm ff}^\varphi - B_{\rm vac}^\varphi} - \beta^r \right)_{R=0}, \label{eq:V2_KS2}
\end{eqnarray}
where we have used equations (\ref{eq:Hff_KS}) and (\ref{eq:Hvac_KS}) to derive equation (\ref{eq:V1_KS2}). Eliminating $D_{\rm ff}^\theta - D_{\rm vac}^\theta$ from equations (\ref{eq:V1_KS2}) and (\ref{eq:V2_KS2}) leads to
\begin{equation}
  V = \frac{\pm \alpha}{\sqrt{\gamma}} \sqrt{\frac{\gamma_{\varphi\varphi}}{\gamma^{\theta\theta}}} \sqrt{1 + \frac{4\pi \sqrt{\gamma} \eta^\theta}{\alpha \gamma_{\varphi\varphi}(B_{\rm ff}^\varphi - B_{\rm vac}^\varphi)}} - \beta^r.
\label{eq:Vmax_KS}
\end{equation}
The sign of the first term in the right-hand side is not determined in this analysis. However, we have assumed $V<0$, and will confirm that the assumption $V<0$ is consistent with the electromagnetic structure.

If $\eta^\theta = 0$, we have
\begin{equation}
  V = \frac{\pm\alpha}{\sqrt{\gamma}} \sqrt{\frac{\gamma_{\varphi\varphi}}{\gamma^{\theta\theta}}} - \beta^r,
  \label{eq:nullv_KS}
\end{equation}
and 
\begin{equation}
  (B_{\rm ff}^\varphi - B_{\rm vac}^\varphi)_{R=0} = \pm \left. \sqrt{\frac{\gamma^{\theta\theta}}{\gamma_{\varphi\varphi}}} (D_\theta^{\rm ff} - D_\theta^{\rm vac}) \right|_{R=0}.
  \label{eq:etazero}
\end{equation}
Substituting $dr = Vdt$ for equation (\ref{eq:metric}), we find
\begin{equation}
  ds^2 = \left(\sqrt{\gamma_{\varphi\varphi}} d\varphi \pm \frac{\alpha \gamma_{r\varphi}}{\sqrt{\gamma \gamma^{\theta\theta}}} dt\right)^2 + \gamma_{\theta\theta} d\theta^2 \ge 0,
\end{equation}
which has to be $ds^2 = 0$, indicating that the four-velocity of the boundary is null. Because this velocity cannot be realized, we can conclude $\eta^\theta \neq 0$. 

The force-free region always satisfies equation (\ref{eq:Bphi_KS}) and has no diverging quantities, and thus the regularity condition at the horizon (equation \ref{eq:regularity_KS}) automatically becomes satisfied after the boundary crosses the horizon.

We have shown that $\eta^\theta > 0$ in the BL coordinates. Since $\sqrt{\gamma} J^\theta$ is the same in the BL and KS coordinates, we have
  \begin{equation}
    \eta^\theta > 0
  \end{equation}
also in the KS coordinates. This is consistent with our assumption that the outward flowing current is concentrated on the equatorial plane (see the texts below equation~\ref{eq:ampere_ff_BL}). By substituting equation (\ref{eq:Vmax_KS}) for equation (\ref{eq:metric}) we find that $\eta^\theta/(B_{\rm ff}^\varphi - B_{\rm vac}^\varphi)|_{R=0} < 0$ is required for the time-like propagation $ds^2 < 0$. Then we have $(B_{\rm ff}^\varphi - B_{\rm vac}^\varphi)|_{R=0} < 0$.

Equation (\ref{eq:V1_KS}) shows that $H_\varphi^{\rm ff}$ is produced by the displacement current and the cross-field current at the boundary. If the former is dominant, $D_{\rm ff}^\theta|_{R=0} > D_{\rm vac}^\theta|_{R=0}$ is realized so that $H_\varphi^{\rm ff} < 0$. This means $l_{\rm ff}|_{R=0} < l_{\rm vac}|_{R=0}$. The production of the electromagnetic AM flux can also be understood by the same equation as equation (\ref{eq:AMconv_BL}) in the BL coordinates. In this case, equation (\ref{eq:V2_KS2}) with $(B^\varphi_{\rm ff} - B^\varphi_{\rm vac})|_{R=0} < 0$ leads to $V+\beta^r < 0$, which means that the minus sign should be taken in equation (\ref{eq:Vmax_KS}).

The production of the Poynting flux can also be understood by equation (\ref{eq:energyconv_BL}) similarly to the case in the BL coordinates. Here the electromagnetic energy densities are
\begin{eqnarray}
  e_{\rm ff} &=& \frac{1}{8\pi} (E^{\rm ff}_\theta D_{\rm ff}^\theta + B_{\rm ff}^\varphi H_\varphi^{\rm ff} + B^r H_r^{\rm ff}), \\
  e_{\rm vac} &=& \frac{1}{8\pi} (E^{\rm vac}_r D_{\rm vac}^r + E^{\rm vac}_\theta D_{\rm vac}^\theta + B^r H_r^{\rm vac}),  
\end{eqnarray}
where we should note that $H_r^{\rm ff} = \alpha B_r^{\rm ff}$ is different from $H_r^{\rm vac} = \alpha B_r^{\rm vac}$, since $B_r = \gamma_{rr} B^r + \gamma_{r\varphi} B^\varphi$. We confirmed that equation (\ref{eq:E0_BL}) satisfies equation (\ref{eq:energyconv_BL}) also in the KS coordinates.

We confirm that the assumption $V<0$ is consistent with the electromagnetic structure which we found. The equation of motion is written down as
\begin{equation}
  \sigma_{\rm m} U^\nu \nabla_\nu U^r + \rho_{\rm mff} U^t_{\rm ff} (V^r_{\rm ff} - V) (U^r_{\rm ff} - U^r) = \left(\frac{\Delta}{\varrho^2} - \frac{\beta^r V}{\alpha^2} \right) \frac{1}{\gamma^{rr}} \sigma D^r|_{R=0} + \frac{\gamma_{\theta\theta}}{\alpha^2 \sqrt{\gamma}} \eta^\theta H_\varphi|_{R=0}.
\end{equation}
We should have $\sigma D^r|_{R=0} < 0, ~ H_\varphi|_{R=0} < 0$ by the same argument as in the BL coordinates, and we have $\eta^\theta >0$. Thus the Lorentz force is in the $-r$ direction while $V<0$, and will overwhelm the gravitational and inertial forces. The inner boundary layer starts with $V<0$ and it is reasonable that it changes its velocity continuously. Then $V>0$ must not be realized because the acceleration cannot be positive around $V \sim 0, ~U^r \sim 0$. That is, the inner boundary layer keeps $V < 0$.

\subsection{Remarks}
\label{sec:discussion}

As shown above, $H_\varphi^{\rm ff}$ and $\Omega_{\rm F}$, or the electromagnetic AM and Poynting fluxes, are created at the inner boundary which propagates towards the horizon. This is a causal mechanism of the flux production as measured in the coordinate basis.

After the inner boundary becomes very close to the horizon in the BL coordinates or it crosses the horizon in the KS coordinates, it does not affect the exterior, and $H_\varphi^{\rm ff}$ and $\Omega_{\rm F}$ are fixed to be consistent with the regularity condition at the horizon (and at infinity). This implies that {\it no source of $H_\varphi^{\rm ff}$ and $\Omega_{\rm F}$ or the AM and Poynting fluxes is required in the steady state.} The steady poloidal currents are just flowing along the field lines without crossing them, and no force is required to drive the currents in the steady state, partly because the force-free plasma is assumed to have no resistivity. This situation is essentially different from that in a steady pulsar wind, in which the electromotive force $\mathbf{V}_\varphi \times \mathbf{B}$ drives the cross-field current in the rotating star, and the fluxes definitely have the electromagnetic sources, i.e. $\nabla \cdot \mathbf{L}_{\rm p} = -(\mathbf{J}_{\rm p} \times \mathbf{B}_{\rm p}) \cdot \mathbf{m}$ and $\nabla \cdot \mathbf{S}_{\rm p} = -\mathbf{E} \cdot \mathbf{J}_{\rm p}$ (see Section~\ref{sec:intro} and TT14).

As a result, we see that the BH loses its rotational energy directly by $\mathbf{S}_{\rm p}$ along the field lines threading the horizon, as described in Figure~\ref{fig:summary}.

The plasma may have a finite resistivity in more realistic BH magnetosphere or in the numerical simulations. In this case a certain force is required to maintain the steady-state currents. K04 and K09 suggest that a weak $\mathbf{D}$ field component {\it parallel to the $\mathbf{B}$ field line} is induced and drives the currents in the steady state without violating the regularity condition significantly.

The inner boundary approaching the horizon as seen in the BL coordinates looks similar to the stretched horizon in the membrane paradigm at first sight. However, they are essentially different. In the membrane paradigm, $H_\varphi^{\rm ff}$ is produced by the fictitious cross-field current flowing on the stretched horizon with Joule dissipation. On the other hand, we have shown that $H_\varphi^{\rm ff}$ is produced not only by the cross-field current but also by the displacement current.

We should note that mechanism of driving the cross-field current on the inner boundary might be different from that near the equatorial plane which is discussed in Section~\ref{sec:equat}. In the latter case, $D^2 > B^2$ can be realized due to the property of the ergosphere, which drives the cross-field current. On the other hand, the mechanism of driving the cross-field current between the force-free and vacuum regions (and its relation to the property of the ergosphere) may not be understood in our toy model, where $\mathbf{D} \cdot \mathbf{B} = 0$ could be violated there. More studies on the plasma physics as measured by the FIDOs would be required.

\section{Conclusion}
\label{sec:conclusion}

We have generically discussed the axisymmetric Kerr BH magnetosphere in which a collisionless plasma satisfies $\mathbf{D} \cdot \mathbf{B} = 0$ (i.e. there is no gap in the plasma region), and clarified the causal production mechanism of the electromagnetic AM and Poynting fluxes (i.e. $H_\varphi$ and $\Omega_{\rm F}$) along the ergospheric field lines crossing the outer light surfaces and the role of the negative energies as measured in the coordinate basis. Our conclusion is the following.

For the field lines threading the equatorial plane, as shown in K04 and TT14, $H_\varphi$ is produced by the cross-field current flowing in the region where $D^2 > B^2$ near the equatorial plane, and $\Omega_{\rm F}$ will be regulated so that the current crossing region is finite. In this paper, we have shown that the particles in that region can have negative AM and negative energy as measured in the coordinate basis by a feedback from the flux production, and shown by using the one-fluid approximation that those particles flow towards the horizon (see Figure~\ref{fig:summary}). Thus BZ process for these field lines appears to be a similar process to the mechanical Penrose process. We have also compared our arguments to the recent MHD numerical simulation results briefly in Section~\ref{sec:MHDnumerical}. 

For the field lines threading the horizon, the structure of the outward electromagnetic AM and Poynting fluxes (or the poloidal currents and the electric potential differences) must not be created by the horizon, but must be a result from phenomena having occurred outside the horizon in the prior coordinate times. To illustrate this concept, we have built a toy model of a time-dependent state in which the force-free plasma injected continuously between the two light surfaces is filling a vacuum (see Figure~\ref{fig:unsteady}). As a result, we have seen that the fluxes are produced by the contributions from the displacement current and the cross-field current at the in-going boundary (see equations~\ref{eq:AMconv_BL} and \ref{eq:energyconv_BL}). In the steady state, the in-going boundary does not affect the force-free region, and the fluxes are maintained without any electromagnetic source (if the resistivity is negligible). $H_\varphi$ and $\Omega_{\rm F}$ are maintained to be consistent with the regularity condition at the horizon and at infinity. The force-free condition is satisfied along the field lines threading the horizon in the steady state, and then conversion of the AM and energy from the particles is negligible. Thus we support the mathematical treatments of \cite{blandford77} and \cite{beskin13} for determining $H_\varphi$ and $\Omega_{\rm F}$ of the steady-state force-free plasma.

We have shown that the concept of the inflow of negative electromagnetic energy along the field lines threading the horizon is not physically essential. The steady outward Poynting flux should be interpreted just as a result of the currents flowing in the plasma with the electric potential differences. The outward Poynting flux at the horizon in the KS coordinates does not violate causality, because the steady fluxes carry no information. The BH loses its rotational energy directly by this outward Poynting flux without being mediated by any infalling negative-energy objects, as described in Figure~\ref{fig:summary}.

Finally, we should emphasize that our analysis is based on several assumptions (see Sections~\ref{subsec:magnetosphere}, \ref{sec:penrose}, and \ref{sec:unsteady}). Our arguments for the particle motions near the equatorial plane and in the toy model are required to be justified by numerical simulations. As for the electromagnetic field, the principal assumption is $\mathbf{D} \cdot \mathbf{B} = 0$ in the steady state. It is still debated whether this condition is satisfied in the steady state at the boundary between the inflow and the outflow (i.e. at $r = r_i$ in our toy model) \citep{beskin00,okamoto06,toma12,broderick15,levinson11,moscibrodzka11} and even in the whole BH (or pulsar) magnetosphere \citep{beskin_rafikov00,petrova15,kojima15}. This issue is closely related to radiation physics, which should be resolved for validating theories on BZ process by observations \citep{asada14,kino15,broderick15}.

\section*{Acknowledgment}

We thank the referee for his/her useful comments. A part of this paper has already been presented by K.T. in the workshop ``Relativistic Jets: Creation, Dynamics, and Internal Physics" held at Krakow, 20-24 April 2015. K.T. thanks its organizers for the wonderful hospitality and its participants for stimulating him to improve the discussion in this paper. K.T. also thanks T. Harada, S. Koide, Y. Kojima, K. Nakao, and Y. Sekiguchi for useful discussions. This work is partly supported by JSPS Grants-in-Aid for Scientific Research 15H05437 and also by JST grant ``Building of Consortia for the Development of Human Resources in Science and Technology.''


%

\appendix

\section{Convective currents and the force-free condition}
\label{app:current}

The relation between convective current and velocity of particles is summarized as follows. Let us consider the case in which the positively and negatively charged particles have the same velocity $\mathbf{v}$ as measured in the coordinate basis. Generalization to other cases is easy. The local physics as measured by the FIDOs indicates
\begin{equation}
\hat{\mathbf{j}} = \rho \hat{\mathbf{v}},
\end{equation}
where we note that $\rho = -I^\mu n_\mu$ is a scalar. Each spatial component in respect of the BL FIDO's orthonormal basis can be rewritten as (TT14)
\begin{equation}
\sqrt{\gamma_{\varphi\varphi}} j^\varphi = \rho\frac{\sqrt{\gamma_{\varphi\varphi}}}{\alpha} (v^\varphi + \beta^\varphi), ~~~
\sqrt{\gamma_{rr}} j^r = \rho \frac{\sqrt{\gamma_{rr}}}{\alpha} v^r.
\end{equation}
(The form of the $\theta$ component is the same as the $r$ component.) Note that $j^\mu = \gamma^{\mu\nu} I_\nu$ is a four-vector, while the particle velocity is $v^i = u^i/u^t$ in terms of the four-velocity $u^\mu$. Then we can write
\begin{equation}
\mathbf{j} = \frac{1}{\alpha} \rho (\mathbf{v} + \boldsymbol{\beta}).
\end{equation}
This relation is valid also in the KS coordinates. Equation (\ref{eq:relation_j}) leads to
\begin{equation}
\mathbf{J} = \rho \mathbf{v}.
\end{equation}

As an example, the velocity of the $\mathbf{D} \times \mathbf{B}$ drift is (TT14)
\begin{equation}
\hat{\mathbf{v}}_{\rm d} = \frac{\hat{\mathbf{D}} \times \hat{\mathbf{B}}}{\hat{B}^2}, ~~~ \mathbf{v}_{\rm d} = \alpha \frac{\mathbf{D} \times \mathbf{B}}{B^2} - \boldsymbol{\beta}.
\end{equation}
Then we have the drift current as
\begin{equation}
\mathbf{j}_{\rm d} = \rho\frac{\mathbf{D} \times \mathbf{B}}{B^2}.
\end{equation}
Under the assumptions in Section~\ref{sec:setup}, the current is generally $\mathbf{j} = \mathbf{j}_{\rm d} + C \mathbf{B}$ for $D^2 < B^2$, where $C$ is a scalar factor. This corresponds to the force-free condition,
\begin{equation}
\rho \mathbf{D} + \mathbf{j} \times \mathbf{B} = 0,
\end{equation}
which is equivalent to $\rho \mathbf{E} + \mathbf{J} \times \mathbf{B} = 0$.

\section{Calculations in the KS coordinates}
\label{app:B2D2}

Here we explain how equations (\ref{eq:B2D2}), (\ref{eq:l_KS}), and (\ref{eq:e_KS}) are derived, and examine the sign of $\mathbf{D} \cdot \mathbf{E}$ in the KS coordinates. The following identities are useful for such calculations:
\begin{eqnarray}
  && \Sigma - 4r^2 = \Delta (\varrho^2 + 2r), \\
  && \Sigma - (\varrho^2 + 2r)a^2 \sin^2\theta = \varrho^4, \label{eq:identity2}\\
  && \Sigma(\varrho^2 - 2r)+ 4r^2 a^2 \sin^2\theta = \varrho^4 \Delta. \label{eq:identity3} 
\end{eqnarray}

From equations (\ref{eq:relation_E}) and (\ref{eq:omegaF}), generally one has
\begin{equation}
  D^2 = \frac{1}{\alpha^2} \left[(\omega^2 + \beta^2 + 2\omega^i \beta_i)B^2 - (\omega^i B_i + \beta^i B_i)^2 \right].
  \label{eq:D2cal}
\end{equation}
In the KS coordinates, this equation is reduced to
\begin{eqnarray}
  (B^2 - D^2)\alpha^2 = - B^2 f(\Omega_{\rm F}, r, \theta) + (\Omega_{\rm F}B_\varphi + \beta^r B_r)^2, 
\end{eqnarray}
where
\begin{equation}
  \Omega_{\rm F}B_\varphi + \beta^r B_r = (\gamma_{r\varphi} \Omega_{\rm F} + \beta_r) B^r + \gamma_{\varphi\varphi} (\Omega_{\rm F} - \Omega) B^\varphi.
\end{equation}
By using equation (\ref{eq:f}) and the above identities, we derive equation (\ref{eq:B2D2}).

The electromagnetic AM density is written by using equations (\ref{eq:relation_E}) and (\ref{eq:omegaF}) as
\begin{eqnarray}
  l = \frac{1}{4\pi} e_{\varphi jk} D^j B^k
  = \frac{-1}{4\pi\alpha} \left[(\omega_i + \beta_i) B^i B_\varphi - (\omega_\varphi + \beta_\varphi) B^i B_i )\right].
  \label{eq:app_l1}
\end{eqnarray}
In the KS coordinates, one has
\begin{equation}
  4\pi \alpha l = -(\omega_r + \beta_r)B^r B_\varphi + (\omega_\varphi + \beta_\varphi)(B^r B_r + B^\theta B_\theta),
  \label{eq:app_l2}
\end{equation}
which can be straightforwardly rewritten as equation (\ref{eq:l_KS}).

The electromagnetic energy density is written by using equations (\ref{eq:relation_E}) and (\ref{eq:relation_H}) as
\begin{eqnarray}
  e = \frac{1}{8\pi} (E_i D^i + B^i H_i)
    = \frac{\alpha}{8\pi} (D^2 + B^2) + \frac{1}{4\pi} D^i e_{ijk} \beta^j B^k.
\end{eqnarray}
Then by using equation (\ref{eq:D2cal}), one has
\begin{equation}
  8\pi \alpha e = (\alpha^2 -\beta^2 + \omega^2)B^2 - (\omega^i B_i)^2
  + (\beta^i B_i)^2,
\end{equation}
which can be straightforwardly rewritten as equations (\ref{eq:e_KS}).

For the field lines threading the equatorial plane, $B^2 - D^2 < 0$ can be realized near that plane, where $B^r = 0$, $H_\varphi = 0$ [which lead to $B^\varphi = 0$ by equation (\ref{eq:Bphi_KS})], and $\Omega_{\rm F} < \Omega$ (TT14). Let us confirm $\mathbf{D} \cdot \mathbf{E} < 0$ in the KS coordinates. Generally this quantity can be calculated by using equations (\ref{eq:app_l1}) and (\ref{eq:app_l2}) as
\begin{eqnarray}
  \mathbf{D} \cdot \mathbf{E} &=& E_i D^i = -e_{ijk} \omega^j B^k D^i
  = \omega^\varphi e_{\varphi jk} D^j B^k \nonumber \\
  &=& \frac{\Omega_{\rm F}}{\alpha} \left[\gamma_{\varphi\varphi} (\Omega_{\rm F} - \Omega) (B^\theta B_\theta + B^r B_r)
    - (\omega_r + \beta_r) B^r B_\varphi \right].
\end{eqnarray}
The conditions $B^r = 0$ and $\Omega_{\rm F} < \Omega$ lead to $\mathbf{D} \cdot \mathbf{E} < 0$. Such a $\mathbf{D}$ field drives the poloidal current to flow in the direction of $-\mathbf{E}$.

\end{document}